\documentclass[12pt,preprint]{aastex}
\def \lp{\>\> .}
\def \lc{\>\> ,}
\def \arcsec{\hbox{$^{\prime\prime}$}}
\def \arcmin{\hbox{$^{\prime}$}}
\def \deg{$^\circ$}

\def \c2{cm$^{-2}$}
\def \cc{cm$^{-3}$}
\def \ccs{cm$^{3}$ s$^{-1}$}
\def \kms{km\ s$^{-1}$}
\def \kks{K kms$^{-1}$}
\def \nh2{n_{H_2}}
\def \nh1{n_{HI}}

\def \nh3{NH$_3$}
\def \n2h{N$_2$H$^+$}
\def \tw{$^{12}$CO}
\def \th{$^{13}$CO}

\def \csev{C$^{17}$O}
\def \hh{H$_2$}
\def \hho{H$_2$O}

\def \oxy{O$_2$}
\def \ox18{$^{16}$O$^{18}$O}
\def \co{CO}

\def \Ms{$M_{\odot}$}
\def \mic{$\mu$m}

\def \be{\begin{equation}}
\def \ee{   \end{equation}}
\def \bf {\begin{figure}}
\def \ef {   \end{figure}}
\def \bc{\begin{center}}
\def \ec{\begin{center}}

\def \lc{\>\> ,}
\def \lp{\>\> .}

\begin{document}
\title{{\it Herschel}\footnote{ {\it Herschel} is an ESA space observatory with science instruments provided by European-led Principal Investigator consortia and with important participation from NASA.} Measurements of Molecular Oxygen in Orion}
\setcounter{footnote}{0}

\author{
Paul F. Goldsmith\altaffilmark{1},
Ren\'{e} Liseau\altaffilmark{2}, 
Tom A. Bell\altaffilmark{3},
John H. Black\altaffilmark{2},
Jo-Hsin Chen\altaffilmark{1},
David Hollenbach\altaffilmark{4},
Michael J. Kaufman\altaffilmark{5},
Di Li\altaffilmark{1},
Dariusz C. Lis\altaffilmark{6}, 
Gary Melnick\altaffilmark{7},
David Neufeld\altaffilmark{8}, 
Laurent Pagani\altaffilmark{9},
Ronald Snell\altaffilmark{10},
Arnold O. Benz\altaffilmark{11},
Edwin Bergin\altaffilmark{12},
Simon Bruderer\altaffilmark{11},
Paola Caselli\altaffilmark{13},
Emmanuel Caux\altaffilmark{14},
Pierre Encrenaz\altaffilmark{9},
Edith Falgarone\altaffilmark{15},
Maryvonne Gerin\altaffilmark{15},
Javier R. Goicoechea\altaffilmark{3},
\AA ke Hjalmarson\altaffilmark{2},
Bengt Larsson\altaffilmark{16},
Jacques Le Bourlot\altaffilmark{17},
Franck Le Petit\altaffilmark{17}
Massimo De Luca\altaffilmark{17},
Zsofia Nagy\altaffilmark{18},
Evelyne Roueff\altaffilmark{17},
Aage Sandqvist\altaffilmark{16},
Floris van der Tak\altaffilmark{18},
Ewine F. van Dishoeck\altaffilmark{19},
Charlotte Vastel\altaffilmark{14},
Serena Viti\altaffilmark{20}, \&
Umut Y{\i}ld{\i}z\altaffilmark{21}
}

\vspace{0.2cm}
\altaffiltext{1}  {Jet Propulsion Laboratory, California Institute of Technology, 4800 Oak Grove Drive, Pasadena CA 91109 Paul.F.Goldsmith@jpl.nasa.gov}
\altaffiltext{2}  {Department of Earth \& Space Sciences, Chalmers University of Technology, Onsala Space Observatory, SE-439 92 Onsala, Sweden}
\altaffiltext{3}  {Centro de Astrobiolog\'{i}a, CSIC--INTA, 28850, Madrid, Spain}
\altaffiltext{4}  {SETI Institute, Mountain View CA 94043}
\altaffiltext{5} {Department of Physics and Astronomy, San Jos\'{e} State University, San Jose CA 95192}
\altaffiltext{6}  {California Institute of Technology, Cahill Center for Astronomy and Astrophysics 301-17, Pasadena CA 91125 }
\altaffiltext{7}  {Harvard-Smithsonian Center for Astrophysics, 60 Garden Street, MS 66, Cambridge MA 02138}
\altaffiltext{8}  {Department of Physics and Astronomy, Johns Hopkins University, 3400 North Charles Street, Baltimore, MD 21218}
\altaffiltext{9}{LERMA \& UMR8112 du CNRS, Observatoire de Paris, 61 Av. de l'Observatoire, 75014, Paris, France}
\altaffiltext{10}  {Department of Astronomy, University of Massachusetts, Amherst MA 01003}
\altaffiltext{11}{Institute of Astronomy, ETH Zurich, Zurich, Switzerland}
\altaffiltext{12} {Department of Astronomy, University of Michigan, 500 Church Street, Ann Arbor MI 48109}
\altaffiltext{13}{School of Physics and Astronomy, University of Leeds, Leeds, UK}
\altaffiltext{14}{Universit\'e de Toulouse; UPS-OMP; IRAP; Toulouse, France \& CNRS; IRAP; 9 Av. Colonel Roche, BP 44346, F-31028 Toulouse Cedex 4, France}
\altaffiltext{15}{LRA/LERMA, CNRS, UMR8112, Observatoire de Paris \& \'{E}cole Normale Sup\'{e}rieure, 24 rue Lhomond, 75231 Paris Cedex 05, France}
\altaffiltext{16}{Stockholm Observatory, Stockholm University, AlbaNova University Center, SE-106 91 Stockholm, Sweden}
\altaffiltext{17}{Observatoire de Paris, LUTH, Paris, France}
\altaffiltext{18}{SRON Netherlands Institute for Space Research, PO Box 800, 9700 AV, and Kapteyn Astronomical Institute, University of Groningen, Groningen, The Netherlands}
\altaffiltext{19}{Leiden Observatory, Leiden University, PO Box 9513, 2300 RA, Leiden, The Netherlands and Max-Planck-Institut f\"{u}r Extraterrestrische Physik, Giessenbachstrasse 1, 85748, Garching, Germany}
\altaffiltext{20}{Department of Physics and Astronomy, University College London, London, UK}
\altaffiltext{21}{Leiden Observatory, Leiden University, PO Box 9513, 2300 RA Leiden, The Netherlands}


\begin{abstract}
We report observations of three rotational transitions of molecular oxygen (\oxy) in emission from the \hh\ Peak 1 position of vibrationally excited molecular hydrogen in Orion.  We observed the 487 GHz, 774 GHz, and 1121 GHz lines using HIFI on the \textit{Herschel Space Observatory}, having velocities of 11 \kms\ to 12 \kms\ and widths of 3 \kms. The beam-averaged column density is N(\oxy) = 6.5$\times$10$^{16}$ cm$^{-2}$, and assuming that the source has an equal beam filling factor for all transitions (beam widths 44, 28, and 19\arcsec), the relative line intensities imply a kinetic temperature between 65 K and 120 K.  The fractional abundance of \oxy\ relative to \hh\ is 0.3 -- 7.3$\times$10$^{-6}$.  The unusual velocity suggests an association with a $\sim$ 5\arcsec\ diameter source, denoted Peak A, the Western Clump, or MF4.  The mass of this source is $\sim$ 10 \Ms\ and the dust temperature is $\geq$ 150 K.  Our preferred explanation of the enhanced \oxy\ abundance is that dust grains in this region are sufficiently warm (T $\geq$ 100 K) to desorb water ice and thus keep a significant fraction of elemental oxygen in the gas phase, with a significant fraction as \oxy.  For this small source, the line ratios require a temperature $\geq$ 180 K.  The inferred  \oxy\ column density $\simeq$ 5$\times$10$^{18}$ cm$^{-2}$ can be produced in Peak A, having N(\hh) $\simeq$ 4$\times$10$^{24}$ cm$^{-2}$.  An alternative mechanism is a low-velocity (10 to 15 \kms) C--shock, which can produce N(\oxy) up to 10$^{17}$ cm$^{-2}$.  

\end{abstract}


\setcounter{footnote}{0}

\section{ INTRODUCTION}
\label{intro}

\subsection{Chemistry of Oxygen in Dense Interstellar Clouds}
\label{oxy_chem}
Oxygen is the third most abundant element in the universe, and its form in different regions is thus of considerable astronomical importance.  
\citet{whittet2010} has recently reexamined the question of oxygen abundance as a function of density in the interstellar medium (ISM), and finds that a large fraction of oxygen is unaccounted for at the higher densities which generally correspond to the molecular ISM.  
Models of gas-phase chemistry that do not treat the condensation of species as icy mantles on grains have predicted molecular oxygen (\oxy) to be abundant in well-shielded regions, with X(\oxy) = N(\oxy)/N(\hh) greater than 10$^{-5}$, and thus the most abundant molecular species after \hh\ and \co, with predicted abundance exceeding that for water \citep{herbst1973, prasad1980, graedel1982, leung1984, langer1989, bergin1995, marechal1997b}.  The conditions that generally (but not in every model) maximize the abundance of \oxy\ are low metal abundance (consistent with depletion on grains), high densities, and sufficient time for oxygen chemistry to come into steady state.  

The gas-phase chemistry of \oxy\ in dense clouds is relatively simple.  Formation of \oxy\ is initiated by the reaction of atomic oxygen with H$_3^+$ to form OH$^+$.    H$_3^+$ is produced by cosmic ray ionization of \hh\ and subsequent reaction of the resulting H$_2^+$ with \hh.  The OH$^+$  reacts with \hh\ to produce H$_2$O$^+$, and by a second reaction, H$_3$O$^+$.  This ion dissociatively recombines with an electron, and the two channels of interest result in \hho\ and OH.  The latter species can react with an oxygen atom to form \oxy.   

Uncertainties in several of the key reaction rates have been reduced by laboratory measurements.  The dissociative recombination of H$_3$O$^+$ has been measured  using heavy-ion storage rings by \citet{vejby1997}, by \citet{jensen2000} and by \citet{neau2000}.  The reaction rate for all three of these measurements is in reasonable agreement in magnitude and temperature dependence, with a large thermal rate coefficient $\alpha$ $\simeq$ 3$\times$10$^{-6}$ cm$^3$s$^{-1}$ at 50 K, and a temperature dependence  $\propto$ $T^{-0.8}$.  These results are consistent with the earlier results by \citet{mul1983} obtained using a different technique.  The branching ratio has been a subject of some uncertainty, the papers cited above all put the fraction of the recombinations that result in production of OH in the range 0.5 $\leq$ $f$(OH) $\leq$ 0.7.  Thus, the effective rate coefficient for formation of OH by dissociative recombination of H$_3$O$^+$  at the temperature of dense molecular clouds is about 2$\times$10$^{-6}$ cm$^3$s$^{-1}$.

The rate at low temperatures of the reaction OH + O $\rightarrow$ \oxy, the final step in the production of molecular oxygen, has been the subject of some controversy.  Theoretical calculations by \citet{harding2000} were in moderately good agreement with the measurements for temperatures between 150  K and 500 K  \citep{howard1981, smith1994}, and indicated the peak value of the rate coefficient to be $k$ = 5.5$\times$10$^{-11}$ \ccs\ at $\simeq$ 80 K, with a factor $\simeq$ 2 reduction at 10 K and 1000 K.  Measurements by \citet{carty2006} sampled a range of temperatures, but these authors recommended use of a single  value, $k$ = 3.5 $\pm$ 1.0$\times$10$^{-11}$ \ccs\ over the temperature range 39 K $\leq$ T $\leq$ 142 K.   

The calculations by \citet{xu2007} suggested a rate that while only somewhat lower than that of \citet{carty2006} at temperatures above 80 K, fell dramatically at lower temperatures.  
This opened the possibility that the \oxy\ abundance in cooler clouds would be much lower than predicted by the above astrochemical models.  However, more elaborate quantum mechanical calculations by \citet{quemener2008} showed a relatively weak temperature dependence, and a rate coefficient at 10 K that is 70 times larger than predicted by \citet{xu2007}.

Recent quantum calculations by \citet{lique2009a} give rate coefficients that are a factor $\simeq$ 2 greater than the measurements of \citet{carty2006} over their temperature range.  The rate coefficients from these latest theoretical calculations do fall at temperatures below 30 K, but even at 10 K are $\geq$ 3$\times$10$^{-11}$ \ccs, thus guaranteeing that the OH + O $\rightarrow$ \oxy\ + H reaction will not be a low-temperature bottleneck for \oxy\ production.  The agreement between measured and theoretical calculations of the reaction rates suggests that the \oxy\ production rate is reasonably well understood and confirms the purely gas-phase chemistry prediction of large X(\oxy) in all but possibly the coldest clouds.  

The expectation of a high abundance of \oxy\ had previously been an impetus to include this species in calculations of gas-phase cooling \citep{goldsmith1978}, and it was found to be moderately important at intermediate densities, despite its weak magnetic dipole transitions.  The major obstacle to studies of astronomical \oxy\ is the strong absorption by this same species in the Earth's atmosphere.  The large scale height of \oxy\ compared to that of \hho\ means that even going to high altitude sites does not allow for observations of Galactic \oxy\ in the interstellar medium, where only the energy levels producing millimeter and submillimeter rotational transitions will likely be populated.  Although absorption observations of UV transitions are possible from space \citep{smith1984},  clouds with sufficient column density to have detectable transitions of \oxy\ are likely to have sufficiently large dust opacities that such observations will prove difficult.  

\subsection{Grain Surface Oxygen}
\label{grain_oxygen}

A possibility that must be considered is that \oxy\ is resident on dust grain surfaces.  Solid \oxy\ has been searched for in interstellar clouds through its weak feature at 6.45 \mic, with negative results \citep{vandenbussche1999}. The most stringent abundance limits come from analysis of the solid \th\  line profiles in low- and high-mass star-forming regions, which indicate solid \oxy\ concentrations of less than 10\% with respect to water ice \citep{boogert2002, pontoppidan2003}.  Taking a typical water ice abundance of 10$^{-4}$ relative to \hh, this indicates solid \oxy/gas-phase \hh\ abundance ratios $\leq$ 10$^{-5}$ within a factor of 2. Thus, at most 10\% of elemental oxygen is locked up in \oxy\ on grain surfaces.

Laboratory measurements show that solid \oxy\ ice is only slightly less volatile than CO ice with a binding energy of $\sim$ 900 K, and thermally evaporates at 16--18 K under typical interstellar conditions \citep{acharyya2007}.  If trapped in water ice, a small fraction of \oxy\ can desorb at higher temperatures together with H$_2$ \citep{collings2004}.  Once on the grains, solid \oxy\ is chemically active.  A number of recent laboratory experiments have demonstrated that \oxy\ is rapidly hydrogenated by atomic H to form OH, H$_2$O$_2$ and H$_2$O in ices, even at temperatures as low as 12 K \citep{miyauchi2008, ioppolo2008, ioppolo2010}.  Some gas-phase \oxy\ can also be produced by photodesorption of H$_2$O ice at high temperatures \citep{oberg2009b}. The nondetection of solid \oxy\ on grain surfaces is thus reasonable, and overall, the amount of oxygen on dust grains is not a major contributor to the overall oxygen budget in dense interstellar clouds.

\subsection{Searches for Molecular Oxygen}

Several different techniques have been utilized in the search for interstellar \oxy.  In general, these normalize the derived column density (or upper limit thereto) to that of standard tracer, generally an isotopologue of carbon monoxide, from which the molecular hydrogen column density can be derived.  The resultant fractional abundance of or upper limit to the fractional abundance of \oxy\ is obtained assuming that the distributions of the species are the same.  The isotopologue \ox18\ has levels and transitions that are not allowed for  the homonuclear common isotopologue, and some of these are at frequencies at which the atmosphere is reasonably transparent, notably the (N,J) = (2,1) $\rightarrow$ (0,1) transition at  233.95 GHz \citep{black1984, marechal1997c}.  Ground-based observational searches for \ox18\  have been carried out with nondetections giving upper limits X(\oxy) somewhat lower than those for CO \citep{goldsmith1985, liszt1985, liseau2010}.  
A ground-based search  \citep{pagani1993} yielded a tentative detection in one source that was not confirmed by further observations \citep{marechal1997a} with significantly improved limits.  
\citet{fuente1993} searched for the 234 GHz transition of \ox18\ in three dark clouds, and measured  1$\sigma$ upper limits N(\oxy)/N(\co) $\leq$ 0.15, 0.19, and 0.29 for  in TMC2, L134N, and B335, respectively.  

\begin{figure}
\begin{center}
\includegraphics[width=9cm] {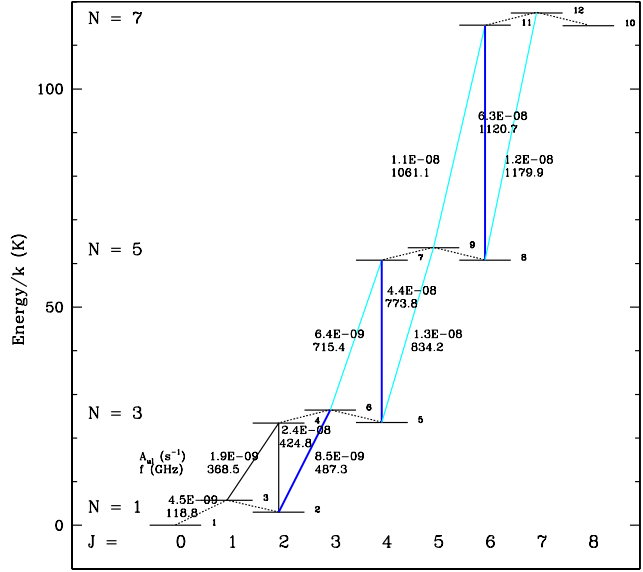}
\caption{Lower rotational energy levels and transitions of \oxy.  Each level is described by its rotational quantum number N and total angular momentum quantum number J, with J = N--1, N, N+1.  This designation applies rigorously only to Hund's case (b), as described by \citet{lique2010}, but is generally used to designate \oxy\ levels.  The frequencies of the allowed transitions are given in GHz, together with the spontaneous decay coefficients in s$^{-1}$.  Transitions with $\Delta$N = 0 are indicated by dotted lines; these have frequencies in the vicinity of 60 GHz with the exception of the N,J = 1,1 to 1,0 transition at a frequency of 118.8 GHz.  The seven transitions that fall in the frequency range covered by {\it Herschel} HIFI are shown in color, with the three relatively strong transitions discussed here (487, 774, and 1121 GHz) are shown in heavy dark blue.  See \citet{marechal1997a} for additional details.}
\label{oxy_levels}
\end{center}
\end{figure}

The common isotopologue of molecular oxygen can be observed using ground-based telescopes from sources having velocities sufficient to Doppler--shift the observed frequency ``out from under'' the pressure-broadened terrestrial \oxy\ absorption.  
The lower rotational transitions of \oxy\ referred to here are shown in Fig. \ref{oxy_levels}.
This technique has been used to search for \oxy\ in the galaxy NGC 7674 by \citet{liszt1985}, using the redshifted 119 GHz line, who found X(\oxy) $\leq$ 1.4$\times$10$^{-5}$ (1 $\sigma$). 
\citet{goldsmith1989} searched for  the 119 GHz \oxy\ transition in the galaxy VII Zw 31, determining a 1 $\sigma$ upper limit X(\oxy) $\leq$ 4$\times$10$^{-6}$.
\citet{combes1991} searched for \oxy\ in NGC  6240 as well as \ox18\ in two Galactic sources, obtaining upper limits X(\oxy) $\leq$ 1$\times$10$^{-6}$. 
\citet{liszt1992} added three galaxies having recession velocities between 8800 and 12,000 \kms\ and carried out additional observations of NGC 7674, finding N(\oxy)/N(CO) $\leq$ 0.05 at the 1$\sigma$ level.
\citet{combes1995} searched for red-shifted 368 GHz and 424 GHz \oxy\ lines in the BL Lac object  B0218+357 having z = 0.685, determining  N(\oxy)/N(\co) $\leq$ 0.014.
\citet{combes1997} observed this same source, but in the 56.3 and 119 GHz \oxy\ transitions, and obtained an improved upper limit N(\oxy)/N(\co) $\leq$ 2$\times$10$^{-3}$.  
\citet{frayer1998} searched for \oxy\ in Markarian 109, obtaining an upper limit N(\oxy)/N(CO) $\leq$ 0.31 in this low-metallicity galaxy.  All of the above ground--based observations had angular resolutions $\leq$ 1\arcmin.

Observations of \oxy\ are possible from balloon altitude for sources within the Milky Way with sufficiently large velocities.  \citet{olofsson1998} searched for the 424 GHz transition of \oxy\ from an altitude of 39.5 km with a 60 cm telescope giving a beam width $\simeq$ 5\arcmin.  They did not detect \oxy, but were able to determine upper limits in two sources, NGC7538 and W51, with N(\oxy)/N(\co) $\leq$ 0.1.  Since X(\co) is expected to be $\simeq$ 10$^{-4}$ in warm regions, these limits correspond to X(\oxy) $\leq$ 10$^{-5}$.  

The best environment for searching for molecular oxygen in the Milky Way as well as other galaxies is clearly above the Earth's atmosphere, from a spacecraft.  Out of a number of proposals, two modest-sized space missions dedicated to submillimeter wavelength spectroscopy have been successfully developed and operated.
The first of these was NASA's \textit{Submillimeter Wave Astronomy Satellite (SWAS)} launched in 1998, which observed the 487 GHz \oxy\ line \citep{melnick2000} with a beam width 3\arcmin.5 x 5\arcmin.  Numerous clouds in the Milky Way were observed, but there were no definitive detections of \oxy.  Upper limits reported by \citet{goldsmith2000} were as low as X(\oxy) $\leq$ 2.6$\times$10$^{-7}$, and a combination of data from 9 sources yielded $\langle$X(\oxy)$\rangle$ = [0.33$\pm$1.6(3$\sigma$)]$\times$10$^{-7}$.  

The {\it Odin} Satellite \citep{nordh2003} was launched in 2001.  It was equipped with a larger antenna and more flexible receiver system than those of {\it SWAS}.  Observations of \oxy\ were primarily of the 119 GHz transition with a beam width of 9\arcmin\ \citep{pagani2003}, giving upper limits to X(\oxy) in half of the sources being as low as 10$^{-7}$.  This survey included relatively warm GMCs, but due to the low energy of the upper level of this transition, the upper limits for \oxy\ in dark clouds were significantly lower than those obtained with {\it SWAS}.  A study of the Sgr A cloud in the Galactic Center yielded a 3$\sigma$ upper limit X(\oxy) $\leq$ 1.2$\times$10$^{-7}$ \citep{sandqvist2008}.  \citet{wilson2005} used {\it Odin} to search for \oxy\ in the Small Magellanic Cloud galaxy, and after 39 hours of integration obtained a 3$\sigma$ upper limit X(\oxy) $\leq$ 1.3$\times$10$^{-6}$ in this source that has a (total) oxygen abundance less than 0.2 of that in the Orion nebula in the Milky Way.

A very deep {\it SWAS} integration on a mosaic of positions in the $\rho$ Ophiuchi cloud yielded a detection of the 487 GHz line at modest statistical significance \citep{goldsmith2002}.  The 5.5 \kms\ velocity of the observed emission line fell within the red-shifted wing of a low-velocity molecular outflow. Since the column density of the outflowing gas is modest, the fractional abundance of \oxy\ derived was quite large, X(\oxy) = 1$\times$10$^{-5}$.  This result was not confirmed by subsequent {\it Odin} observations of the 119 GHz line, which did detect an emission feature, but at a velocity of 3.5 \kms, and linewidth of 1.5 \kms, corresponding to the characteristics of the quiescent cloud rather than the outflow.  The line is quite weak and due to the large column density of the bulk of the source, the derived fractional abundance of \oxy, X(\oxy) = 5$\times$10$^{-8}$ \citep{larsson2007}, is relatively low.   The consensus emerging from these observations of \oxy\ is that the abundance of this species in a variety of different types of molecular clouds is dramatically lower than predictions of standard gas-phase chemistry.  Instead of an abundance X(\oxy) $\simeq$ 10$^{-5}$, the upper limits are as low as 10$^{-7}$, only slightly above the level of the {\it Odin} $\rho$ Ophiuchi detection.  

However, even preceding the arrival of data from space missions, there had been theoretical models of the ISM that predicted lower \oxy\ abundances.  \citet{xie1995} investigated turbulent diffusion, which in principle could accelerate mixing of material from UV-irradiated cloud edges with the material deeper inside.  The result is reduction in the abundances of certain molecules including \oxy.  A model employing circulation has also been developed by \citet{chieze1989}, which can result in a significant reduction in the abundance of \oxy\ if the characteristic circulation time is as short as 10$^6$ yr.  \citet{lebourlot1995} studied a bistability in interstellar chemistry that under certain conditions can result in a high ionization phase in which the abundance of \oxy\ is reduced by a factor $\simeq$ 100 relative to standard gas-phase chemistry.  This is largely restricted to regions of  low-to-moderate density, but can be favored by high cosmic ray fluxes.  

Results from {\it SWAS} and {\it Odin} indicated that the low abundance of \oxy\ is widespread, and included very well-shielded regions of high density with n(\hh) $\geq$ 10$^4$ \cc.
Additionally, the abundance of \hho\ was found to be substantially below that expected from gas-phase chemistry models.  To address this,   \citet{bergin2000} developed a model invoking depletion of atomic oxygen onto dust grains, and conversion to water ice remaining on the grain surface.  This model results in  an increase of the gas-phase C/O ratio, which has the effect of dramatically reducing the fractional abundance of \oxy\ deep inside clouds.  \citet{hollenbach2009} later showed that at intermediate depths ($A_V \sim 2-8$), photodesorption of water ice from grains maintained a sufficiently high gas-phase abundance of elemental oxygen that abundances X(\oxy) $\sim$ 3$\times$10$^{-7}$ were achieved in these regions, independent of the gas density or incident ultraviolet flux.  

The situation regarding \oxy\ has remained sufficiently puzzling that an Open Time Key Program (Herschel Oxygen Project, or HOP) was proposed and selected to carry out a survey of warm molecular clouds in three rotational transitions of this elusive species.  
This paper presents the results for some of the observations in Orion, which we give in \S \ref{observations}.
In \S \ref{identification}, we discuss the identification as \oxy.
In \S \ref{analysis}  we discuss collisional excitation of \oxy, the source temperature and \oxy\ column density derived assuming the source fills all HIFI beams. 
In \S \ref{filling_offset_effects} we analyze the effects of beam filling and source offset, and discuss the column density and fractional abundance of \oxy\ that apply to different possible source geometries.  
We discuss shocks and warmed grains as sources of the observed \oxy\ in \S \ref{discussion}.
In \S \ref{conclusions} we present a summary of our results and conclusions.

\section{ HIFI Observations}
\label{observations}

\subsection{Data Taking}
\label{data_taking} 
These data were taken with the Heterodyne Instrument for the Far Infrared (HIFI) \citep{degraauw2010}\ on the \textit{Herschel Space Observatory} \citep{pilbratt2010}.  The observations were carried out on operational days (OD) 291, 297, 474, 504, and 505 in March, August, September, and  October 2010.  The position observed is the direction of the NW maximum of \hh\ ro--vibrational emission \citep{beckwith1978}.  The coordinates are $\alpha_{J2000}$ =  5$^h$35$^m$13$^s$.7, $\delta_{J2000}$ = -05\deg22\arcmin09\arcsec.   Double beam switching was employed.  For observations of Orion the spacecraft pointing constraints meant that the reference positions were located within 10$^o$ of an east--west line, with one position being 3\arcmin\  east of the source position and the other being symmetrically located to the west.  Eight settings of the local oscillator were used to enable sideband deconvolution.  HIFI band 1a was used for the 487 GHz observations, band 2b for the 774 GHz observations, and band 5a for the 1121 GHz observations.    Observations of the J = 7 -- 6 \csev\ transition were obtained simultaneously with the 774 GHz \oxy\ data, and the results are discussed in \S\ref{velloc}.

\begin{figure}
\begin{center}
\includegraphics[angle = 270, width = 9cm]{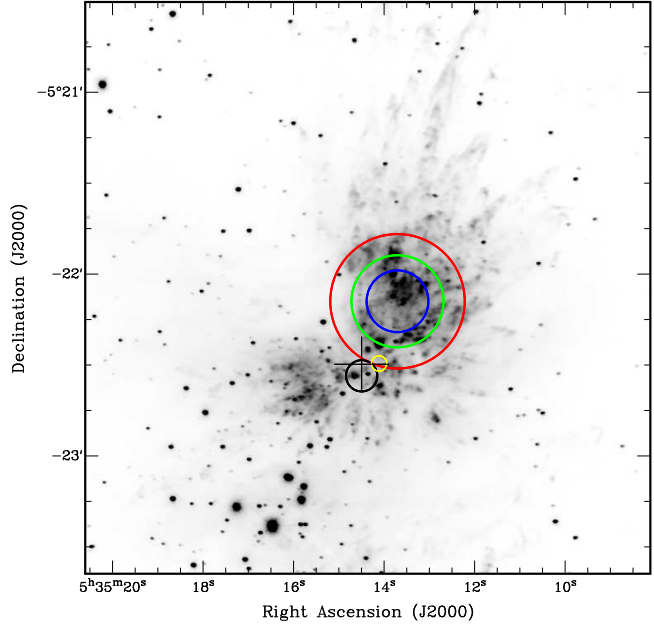}
\caption{Image of vibrationally excited 2 \mic\ \hh\ emission in Orion \citep[from] []{bally2011}.  The coordinates of  \hh\ Peak 1 towards which {\it Herschel} was pointed for these observations are $\alpha_{J2000}$ = 5$^h$35$^m$13$^s$.7,  $\delta_{J2000}$ = $-$5\deg22\arcmin09\arcsec.  The three circles  indicate the FWHM beam sizes at the three observed frequencies:  44\arcsec\ at 497 GHz, 28\arcsec\ at 774 GHz, and 19\arcsec\ at 1121 GHz.  The position of the Hot Core ($\alpha_{J2000}$ = 5$^h$35$^m$14$^s$.5, $\delta_{J2000}$ = $-$5\deg22\arcmin33.6\arcsec) is indicated by the black circle 10\arcsec\ in diameter, approximately the size of this region \citep{wilson2000}. The Peak A - Western Clump - MF4 - Cnt D source (see references in text) is indicated by the yellow circle, 5\arcsec\ in diameter.  The infrared source IRc2 \citep{downes1981} is indicated by the black cross.}
\label{orion_hifi_beams} 
\end{center}
\end{figure}

The total (on--source + off--source) integration time for the 487 GHz observations was 440 minutes for H polarization and 398 minutes for the V polarization, due to the failure of one segment of the observations for the latter.  The integration times were 430 and 409 minutes for the 774 GHz and 1121 GHz observations, respectively.  The FWHM beam width is  44\arcsec\ at 487 GHz, 28\arcsec\ at 774 GHz, and 19\arcsec\ at 1121 GHz, while the main beam efficiency, $\epsilon_{MB}$, is equal to 0.76, 0.75, and 0.64 at the three frequencies, respectively \citep{olberg2010}.  
These long integrations provided a test of the performance of the HIFI radiometers.  As discussed in Appendix 1, the noise in the final spectra is a factor $\sim$ 1.6 greater than  expected theoretically.
Figure \ref{orion_hifi_beams} shows the HIFI beams on an image of the \hh\ emission in Orion, and indicates some of the compact sources in the vicinity.

\subsection{Data Processing}
\label{dataproc}

\begin{figure}
\begin{center}
\includegraphics[width = 9cm]{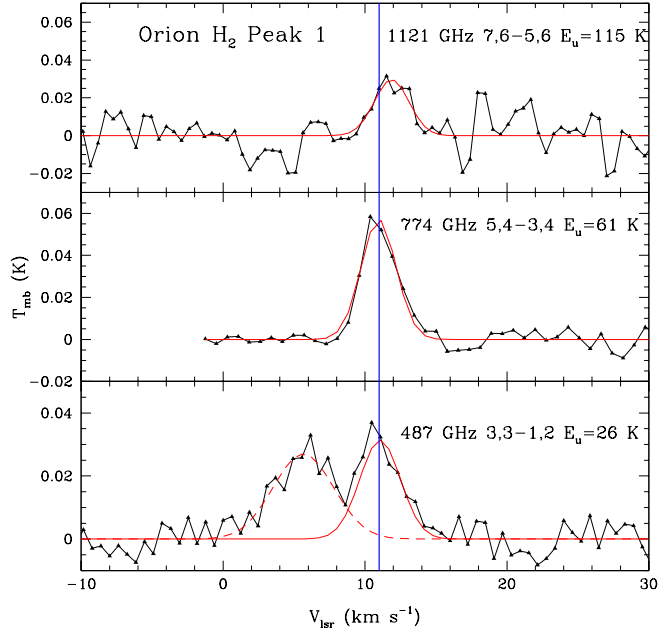}
\caption{\label{o2_3trans} Spectra of the three rotational transitions of \oxy\ observed with the {\it Herschel} HIFI instrument towards the Orion  \hh\ Peak 1 position.  The intensity scale is the main beam brightness temperature (antenna temperature divided by main beam efficiency).  The red solid lines are fitted Gaussians, which all peak at LSR velocities between 11 and 12 \kms.  For the 487 GHz transition there is clearly a second feature, indicated by the broken red line, which we attribute primarily to \oxy\ emission from the Hot Core at v$_{lsr}$ = 5.5 \kms, as discussed in the text.}
\end{center}
\end{figure}

The data were processed using the standard HIFI pipeline in HIPE 4.10 \citep{ott2010}, then exported into FITS format for further analysis using the IRAM Gildas software package (http://www.iram.fr/IRAMFR/GILDAS).  The HIPE task FitHifiFringe was used as part of the pipeline processing to remove standing waves from the data. This was followed by the task DoDeconvolution which utilized the multiple local oscillator settings to extract single sideband spectra.  In Figure \ref{o2_3trans} we show data for a  limited range of  velocities around the frequency expected for each of the lines.  The continuum as well as the wings of nearby strong spectral lines have been removed by baseline fitting.  A broader view of these extremely sensitive submillimeter spectra will be presented elsewhere.

The two orthogonal linear polarizations of each of the HIFI bands are received by separate feedhorns and processed independently.  They thus inevitably have slight relative pointing offset on the sky, which  due to the complex nature of the Orion region,  gives us useful information on the source location, which is discussed in \S\ref{pointing}.  

We see that there is a feature for each transition that peaks between 11 and 12 \kms (all velocities refer to the Galactic Local Standard of Rest).  The results from fitting a single Gaussian to each of these features are given in Table \ref{oxy_param}.  The presence of numerous other lines, especially close by to the 487 GHz \oxy\ line means that the uncertainties in the line parameters are increased by the uncertainties in the fitted baselines.  For the lowest-frequency transition, this may well be the dominant contribution to the total error, so that the uncertainties given in Table \ref{oxy_param} must be regarded as conservative.  The 487 GHz transition shows a second component near 5 \kms, which is a separate entry in Table \ref{oxy_param}.  We attribute this feature primarily to \oxy\ emission from the Hot Core being picked up only in the large beam at this relatively low frequency.  This is the only transition in which there is evidence for this second velocity component.  The issue of the \oxy\ signal coming from the Hot Core, as well as other possible contributions, is discussed further below.

\begin{deluxetable}{cccccc}
\small
\tablewidth{0pt}
\tablecaption{\label{oxy_param} Parameters of Observed \oxy\ Lines}
\tablehead{  \colhead{Transition}  &\colhead{ Frequency}    & \colhead {$\int {\rm T_{mb}dv}$}         &\colhead{V$_0$}                     & \colhead{$\Delta$V (FWHM)} & \colhead{T$_{\rm mb}$ (peak)}\\
                       \colhead{}                    &\colhead{(GHz)}             & \colhead {(K km s$^{-1}$)}          &\colhead{(km s$^{-1}$)}         & \colhead {(km s$^{-1}$)}           & \colhead{(K)}
                       }
\startdata
$3,3$ -- $1,2$    & \,\,\,487.249    & $0.095 \pm 0.011$   & 10.96 $\pm\ 0.16$        & $3.05 \pm 0.40$   & 0.029 \\
                                &                          & $0.113 \pm 0.013$   & \,\,\,5.70   $\pm\ 0.23$   & $4.43 \pm 0.65$   & 0.024 \\
$5,4$ -- $3,4$    & \,\,\,773.840    & $0.177 \pm 0.008$   & 10.96 $\pm\ 0.07$        & $2.91 \pm 0.16$   & 0.057 \\
$7,6$ -- $5,6$    &1120.715        & $0.091 \pm 0.018$   & 11.87 $\pm\ 0.29$         & $2.87 \pm 0.62$   &0.030 \\

\enddata

\end{deluxetable}

\section{ Identification as Molecular Oxygen}
\label{identification}

Given that the integrated intensities represent 8.5 $\sigma$, 22 $\sigma$, and 5 $\sigma$ statistical features, there is little doubt that we have detected emission lines at the frequencies of each of the three \oxy\ transitions that we have observed at the \hh\ Peak 1 position.  The question of whether the features are due to molecular oxygen depends on the agreement and plausibility of the velocities, the consistency of the line intensity ratios, and the presence or absence of other carriers at the three frequencies.  

The  $\simeq$ 11 \kms\ component of the 487 GHz transition and the two higher frequency transitions are all consistent in terms of line width.  The central velocity of the 1121 GHz line is higher by about 0.9 \kms\  than the central velocities of the other two lines.  The formal uncertainty on the fitted central velocity is 0.3 \kms, and the difference in velocities is thus three times this, although less than 1/3 of the FWHM line width.  It is thus possible that, assuming all three lines are from \oxy,  that the velocities are in agreement but we are seeing a statistical fluctuation in the fitting of the central velocity, most likely of the highest frequency line.  Alternatively the highest frequency line is sampling different gas.  

\subsection{\label{interlopers}Interloping Lines}

It seems improbable that a single molecular species other than \oxy\ would have three spectral lines that match so closely with those from the species of interest.  It is more plausible that the three spectral features in question are produced by three different ``interloper'' molecules, but in this case they additionally must have essentially identical line widths.  While we cannot rule out this possibility, it should be born in mind that in this region of Orion, there is a wide variety of line shapes and widths due to the presence of multiple kinematic structures, so the species responsible would have all to share one specific set of line profiles and widths, further reducing the likelihood of this explanation.

We have carried out a targeted analysis to assess likelihood that the observed spectral features are due to known molecular species.  To do this, we first selected all molecular transitions found in the SPLATALOGUE catalog (http://www.splatalogue.net) falling within 5 \kms\ of any of the three \oxy\ lines observed, and having upper state energies less than 1000 K.  There were two species (CH$_3$NH$_2$ and CH$_3$OCHO) having lines possibly interfering with the 487 GHz \oxy\ line, three species ((CH$_3$)$_2$CO, g-CH$_3$CH$_2$OH, and OS$^{18}$O) close to the 774 GHz line, and 6 species (CH$_3$$^{13}$CH$_2$CN, g-CH$_3$CH$_2$OH, HCCCHO, CH$_3$NH$_2$, g-C$_5$H, and t-CH$_2$CHCHO) proximate to the 1121 GHz line.  

For each species, we used the CASSIS program (http://cassis.cesr.fr) to simulate the emission spectrum for a 5\arcsec\ diameter source having line width between 2 \kms\ and 4 \kms, assuming LTE with kinetic temperatures between 75 K and 150 K.  In every case except one (discussed below) we found that we had sufficient frequency coverage to include undetected stronger lines located somewhere in the HIFI spectrum.  From the nondetections, we could set an upper limit to the column density of the species in question that was so low that all lines in the vicinity of the \oxy\ transitions were significantly weaker than the observed \oxy\ main beam brightness  temperatures.  The range of frequency covered in the HIFI spectra ensured that transitions from states of comparable energy above the ground state to those producing possibly interfering lines were observed.  Thus, the gas temperature could not be adjusted in a manner to make the lines close to \oxy\ relatively more intense than transitions of other carriers that were not detected.  An example of this is (CH$_3$)$_2$CO (acetone).  This molecule has a transition only 12 MHz from the 773.8397 GHz \oxy\ line, but also has a group of transitions near 774.4 GHz predicted to be an order of magnitude more intense than the line close to \oxy.  We do not detect the higher frequency lines.  The sensitivity of our data is such that the possible contribution to the \oxy\ line is more than an order of magnitude smaller than the observed feature.  In this manner, we were able to eliminate 11 out of the total of 12 potentially interfering molecular species from consideration.  

One problematic species  is CH$_3$OCHO (methyl formate), whose 40$_{6~34}$ -- 39$_{6~33}$ transition in the v = 1 state of the E species has rest frequency  487251.99996 MHz, compared to 487249.3755 MHz for the 3,1 -- 3,2 \oxy\ transition.  This is the only \oxy\ transition with possible interference from this species.  The upper level energy is 705 K.  A number of narrow features in the full HIFI spectrum are reproduced assuming that methyl formate  is present emitting with a centroid velocity of 8 \kms\ and a line width of 2.5 \kms.  If we assume that we are directly pointed at a 5\arcsec\ source having kinetic temperature 100 K, we derive a column density of 1.5$\times$10$^{17}$ cm$^{-2}$.  The exact value of the column density depends on the location and characteristics of the source relative to the HIFI beams, as well as the complex characteristics of emission from this species  found by \citet{favre2011}.  If the emission is entirely from the MF4 source (see Figure \ref{orion_hifi_beams}), the column density required would be considerably greater than found by \cite{favre2011} so there may well be an extended emission component contributing.  With these values defining the methyl formate emission, a significant portion of the 5.5 \kms\ feature of the 487 GHz \oxy\ line seen in Figure \ref{o2_3trans} could be due to this species.  Note that to reproduce the 11 \kms\ \oxy\ feature, the CH$_3$OCHO would have to be at 12.6 \kms, a value that is outside the range found by \citet{favre2011}, seen in our {\it Herschel} data,  or found for any of the other species in this region of Orion.  

The 5.5 \kms\ feature seen in the 487 GHz \oxy\ transition in the bottom panel of Figure \ref{o2_3trans} is at the velocity observed for most emission from the Hot Core region near KL.  If all or part of this feature is due to \oxy, the fact that we see it only in the 487 GHz spectrum is not likely to be a result of excitation; as discussed at length in \S\ref{excitation}, the \oxy\ transitions are likely to be close to thermalization and the relative intensity of the 774 GHz line compared to that of the 487 GHz line will not be very different from unity.  We are more likely seeing a result of  the very different beam sizes and the offset between the Hot Core and \hh\ Peak 1.  The angular offset is 27\arcsec\, and the 1/e beam radii are 26.3\arcsec, 16.6\arcsec,  and 11.4\arcsec, at 487, 774, and 1121 GHz, respectively, yielding coupling factors to a point source 0.34, 0.07, and 0.003, respectively.  The finite size of the Hot core will modify these results slightly, but it is still the case that when pointed at \hh\ Peak 1, the 487 GHz beam will couple moderately well to the Hot Core, but the two higher frequency beams will have very small coupling, explaining why the 5.5 \kms\ feature is seen only in the lowest-frequency \oxy\ line.  This issue is discussed further in \S \ref{filling_offset_effects}.  

\begin{figure}
\begin{center}
\includegraphics[width = 9cm, angle = 0]{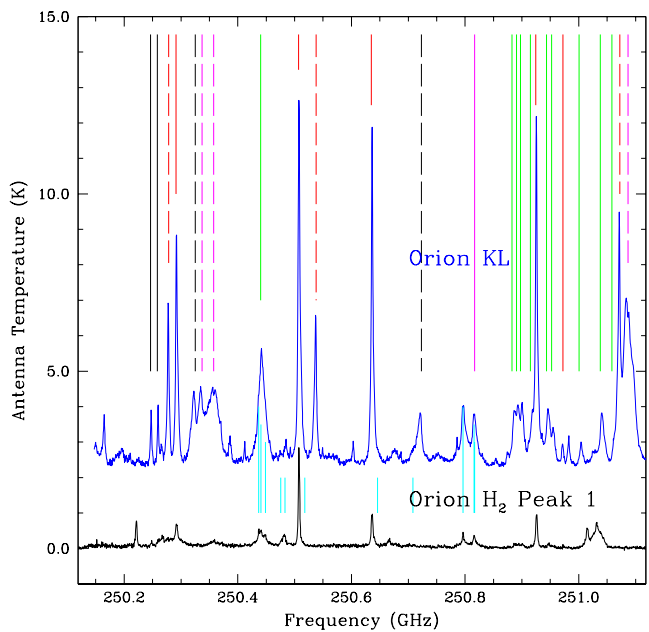}
\caption{Spectra taken at the CSO in January, 2011, while observing the NO molecule.  The upper spectrum is of KL, while the lower spectrum is of \hh\ Peak 1.  The components of the J = 5/2 to J = 3/2 transition of NO are indicated by cyan lines, transitions of CH$_3$OH by red lines, of C$_2$H$_5$CN by green lines, of SO$_2$ and $^{34}$SO$_2$ by magenta lines, of NS by black lines, and of CH$_3$OCHO by black lines.  Identifications of transitions in the signal sideband are indicated by solid lines, and of transitions in the image sideband by dashed lines.  The source velocity has been assumed to be 9 \kms, so that transitions from the hot core, typically having velocities 5 \kms\ to 6 \kms\ are slightly shifted from their identifying lines.
}
\label{NO_wide}
\end{center}
\end{figure}

Another aspect of the confirmation that the three lines shown in Figure \ref{o2_3trans} are indeed produced by \oxy\ is the ``chemical poverty'' of the \hh\  Peak 1 position relative to that of KL and the Hot Core.  This is illustrated by comparing the two spectra in Figure \ref{NO_wide}, which shows a dramatic reduction in the strength of almost all lines at \hh\ Peak 1 compared to KL.  Many species drop by more than an order of magnitude in integrated intensity.  The line crowding in Orion KL is one of the reasons we chose to avoid this source in our initital observations of \oxy\ with HIFI.  A small number of features are stronger, or at least more readily visible at \hh\ Peak 1.  At this point, we have not  been able to make secure identification of most of these.  The conclusion is that finding an ``interloping species'' mimicking the three putative \oxy\ lines in Figure \ref{o2_3trans} is definitely much less likely at \hh\ Peak 1 than it would be at Orion KL. 

\subsection{\label{velloc}\oxy\ Line Velocities and Source Identification}

Studies of the central portion of the Orion Molecular Cloud in different molecular tracers have resulted in a number of compact regions being associated with the velocity range 10 -- 12 \kms, despite this being quite distinct from the nominal velocity of the Hot Core (5 -- 6 \kms) and of the Compact Ridge including KL itself (8 -- 10 \kms).  In what follows, we briefly discuss these sources, with our goal being to show that associating the observed velocity range of the \oxy\ lines with one of these regions is not unreasonable.

\citet{pauls1983} identified a source which they denoted ``Peak A"  in NH$_3$ (3,3) emission, at location $\alpha_{B1950}$ = 5$^h$32$^m$46$^s$.7, $\delta_{B1950}$ --05$^o$24\arcmin24\arcsec.  Although the centroid velocity of this source is 2 \kms, the centroid of the spectrum at offset position (-2\arcsec, 2\arcsec), thus displaced toward our \hh\ Peak 1 position, is $\simeq$ 10 \kms.   
\citet{wilson2000} discuss a 10 -- 12 \kms\ component of the (4,4) transition of NH$_3$.  These authors provide little information about its location, but it is apparently the northwest portion of the Hot core and could be an extension of Peak A.  As such it is likely within 5\arcsec\ of the Hot Core nominal position, offset by $\simeq$ 27\arcsec\ from our \hh\  Peak 1 position.  
\citet{plambeck1987} see a relatively distinct peak in HDO  in a velocity bin spanning 10.5 to 15.5 \kms\ at a position coincident with Peak A, which is 8\arcsec\ to the northwest of the Hot Core, and thus about 21\arcsec\ from \hh\ Peak 1.  \citet{masson1988} present a marginally spectrally distinct feature in the 12 to 14 \kms\ range in HC$_3$N  J = 12 -- 11, in the overall spectrum of the Hot Core, and find that emission at around 13 \kms\ peaks in the ``Western Clump" which is $\simeq$ 5\arcsec\ west of IRc2, and essentially coincident with Peak A.  However, the velocity range 9.5 \kms\ to 11.5 \kms\ is also characteristic of the northern ridge cloud, observed in a number of species by \citet{wright1996}.  

\citet{murata1991} carried out a study of CS J = 2 -- 1 and J = 3 -- 2, and found a number of interesting structures, including numerous clumps which are quite prominent in the velocity range 9.9 -- 11.4 \kms\ in their interferometric map of the higher frequency line.  Most of these lie along a northeast -- southwest line defining the compact ridge, but there are three additional (unnumbered) peaks.  The first is just 2\arcsec\ southeast of \hh\ Peak 1, the second  is about 16\arcsec\ south of our pointing position, and the third is 7.5\arcsec\ east and 20.6\arcsec\ south of our \hh\ Peak 1 position.  This last peak is coincident with the Western Clump -- Peak A feature.  While no details are given about these peaks, they seem to be part of an arc-shaped ring of emission that lies just outside the blue lobe of the CO bipolar outflow mapped by \citet{erickson1982}.  
\citet{murata1992} reported 3 mm continuum mapping of the central portion of Orion, in which a number of condensations were identified.  Of greatest relevance to the present study are the ``Cnt A" peak which coincides with the second CS peak of \citet{murata1991}, and ``Cnt D", which is coincident with the third CS peak, and thus with Western Clump -- Peak A. 

A recent high-angular resolution study of methyl formate (CH$_3$OCHO) by \citet{favre2011} indicates that at least two small clumps have a second kinematic component at velocities higher than the 7 \kms\ to 8 \kms\ characteristic of most of the emission in the center of the Orion molecular cloud surrounding the Hot Core and the Compact Ridge.   One of these clumps, denoted MF4, has its second velocity component, in the three different methyl formate lines observed, at between 10.2 and 11.4 \kms, thus quite similar to the velocities of the \oxy\ lines.     This condensation agrees in position with the Western Clump -- Peak A feature, confirming that this velocity range does seem to be characteristic of this compact source.    The issue of the identification of a specific condensation with the \oxy\ emission will be discussed further in what follows and in \S \ref{discussion}.  

\begin{figure}
\begin{center}
\includegraphics[width = 9cm, angle = 0]{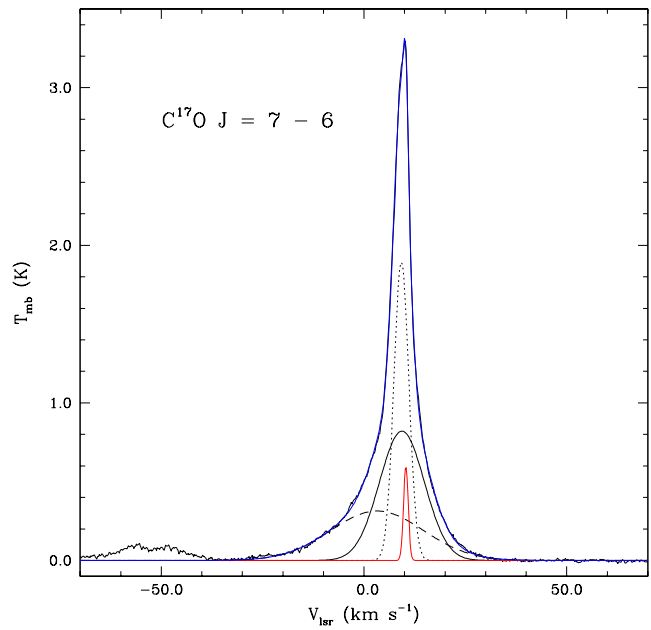}
\caption{ Spectrum of \csev\ J = 7 -- 6 emission at  the \hh\ Peak 1 position obtained  simultaneously with the \oxy\ 774 GHz data using {\it Herschel} HIFI.  The four-Gaussian fit indicates a pedestal feature $\simeq$ 22 \kms\ wide (long-dashed curve) and three narrower components.  The narrowest of these components peaks at a velocity of 10.3 \kms.}
\label{C17O_7-6}    
\end{center}
\end{figure}

One line observed with {\it Herschel} HIFI simultaneously with the 774 GHz \oxy\ line is the J = 7 -- 6 transition of \csev.  A hint of a peak at velocity greater than 10 \kms\  can be seen in the \tw\ J = 7 -- 6  line observed by \citet{pardo2005}, but in the vicinity of KL the profiles are dominated by the broad wings of the outflow.  In Figure \ref{C17O_7-6} we show the spectrum of the line of the rare isotopologue observed with {\it Herschel} HIFI.  The asymmetry in the narrow component of the emission is immediately apparent, and we also show a four-Gaussian component fit, which is the minimum required to reproduce the observed line profile reasonably well. We give the parameters of the components in Table \ref{C17O_7-6_param}.  The presence of the wide component and the asymmetric narrow component leaves the uniqueness of the very narrow narrow component uncertain.  It is clear, however, that there must be a component of the gas having central velocity in excess of 10 \kms\ at the \hh\ Peak 1 position.  In \S  \ref{xo2} we use the \csev\ emission to determine the column density of \hh\ and the fractional abundance of \oxy.

\begin{center}
\begin{deluxetable}{ccccc}
\small
\tablewidth{0pt}
\tablecaption{\label{C17O_7-6_param} Parameters of Observed \csev\ J = 7 -- 6 Line at \hh\ Peak 1}
\tablehead{  \colhead{Compo-}   & \colhead {$\int T_{mb}dv$}         &\colhead{$V_0$}                     & \colhead{$\Delta V$ (FWHM)} & \colhead{T$_{mb}$ (peak)}\\
                       \colhead{nent}                        & \colhead {(K km s$^{-1}$)}          &\colhead{(km s$^{-1}$)}         & \colhead {(km s$^{-1}$)}           & \colhead{(K)}
                       }
\startdata
1                 & \,\,\,9.00    &\,\,\,3.22  &27.0       &0.31\\
2                 &11.40  & \,\,\,9.36 &13.1       &0.82\\
3                 & \,\,\,8.72   & \,\,\,9.21   & \,\,\,4.3       &1.90\\
4	          & \,\,\,0.88   &10.33 & \,\,\,1.4       &0.59\\

\enddata
\end{deluxetable}
\end{center}

\begin{figure}
\begin{center}
\includegraphics[width = 9cm, angle = 0]{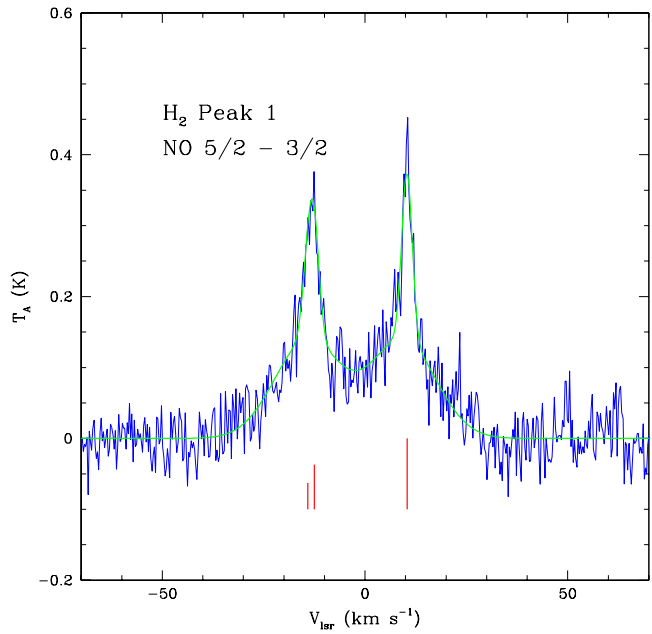}
\caption{Spectrum of the F = 7/2 -- 5/2, F = 5/2 -- 3/2, and F = 3/2 -- 3/2 hyperfine components of the J = 5/2 -- 3/2 transition of NO.  The relative intensities and offsets referenced to the best--fit velocity of 10.4 \kms\ for the low dispersion portion of the strongest F = 7/2 -- 5/2 hyperfine component are indicated by the vertical lines.  The combined best-fit spectrum of the two kinematic components for each of the three hyperfine components is indicated by the solid green curve.
}
\label{NO_narrow}
\end{center}
\end{figure}

Another relevant molecule is NO, which can be produced by reaction of atomic nitrogen with OH, analogous to the pathway for producing \oxy.  We have observed the complete set of the fine-- and hyperfine--structure components of the $^2\Pi_{1/2}\: J = 5/2\to 3/2$  transition, which fall between 249.6 and 251.6 GHz, with the Caltech Submillimeter Observatory (CSO) in January 2011.  The beam size was 32\arcsec\ FWHM and the beam efficiency was 0.73.  The full spectrum is shown in Figure \ref{NO_wide}.  We have focused on the relatively unblended and unconfused trio of hyperfine components near 250.8 GHz, and used CLASS software to fit the spectrum based on optically thin relative line intensities and two line components.  The parameters for the wide component are v$_0$ = 9.3 \kms, $\Delta v_{\rm FWHM}$ = 19.0 \kms, and for the narrow component, v$_0$  = 10.4 \kms, and $\Delta v_{\rm FWHM}$ = 3.1 \kms.  The data and the best-fit profile with these two components are shown in Figure \ref{NO_narrow}.  The excellent fit confirms that there is a component having centroid velocity and line width quite different from those characteristic of the Hot Core or the region around KL, and which are close to the parameters of the \oxy\ lines.

The integrated intensity in the narrow component of the NO 250.796 GHz line is $\int T_{\rm MB} dv = 1.4$ K km s$^{-1}$. Non-LTE models of NO were carried out using the RADEX code and collision rates from \citet{lique2009b}.  They indicate beam-averaged column densities of $2.7\times 10^{15}$ and $1.45\times 10^{15}$ cm$^{-2}$ for $n({\rm H}_2)=5\times 10^5$ cm$^{-3}$, $T_{\rm k}=77.5$ K, and $10^4$ cm$^{-3}$, 100 K, respectively. The corresponding column densities of O$_2$ in a non-LTE treatment are 6.5 and 7.5 $\times 10^{16}$ cm$^{-2}$, respectively. Apparent beam-averaged abundance ratios NO/\oxy = 0.04 and 0.02 are deduced. The excitation of NO is more sensitive to density and temperature than that of O$_2$.  Observations of higher transitions of NO will help to constrain the conditions in the \oxy-emitting region and ultimately could help determine the ratio of atomic nitrogen and oxygen in the gas phase.

HIFI observations of the 620.710 GHz $5_{32}$ -- $4_{41}$ transition of ortho--\hho\ in Orion have yielded, in addition to a broad thermal emission feature seen throughout the central portion of the cloud, a very narrow feature close to 12 \kms (Melnick et al. 2011, in preparation).  This has many characteristics of maser emission, with extremely narrow lineswidths $\leq$ 1 \kms.  The size of the narrow-line emission region is difficult to determine due to the presence of large gradients in the thermal emission, but the source location is likely  coincident with or slightly north of the \hh\ Peak 1 position.  Populating the upper level of this transition, more than 730 K above the ground state (as well as the location of the emission) is more consistent with spatially-extended shocked gas than with any of the compact sources discussed above.

\subsection{\label{pointing} Constraints on Source Location From Pointing Offsets}

The two linear polarizations (H and V) observed by HIFI have a small relative pointing offset, which is different for each band.  There are clear detections in each polarization of the 774 GHz \oxy\ transition each of two well-separated observing epochs (March and September 2010).    The 1121 GHz line lacks sufficient signal to noise ratio to allow epoch- and polarization-dependent analysis, and the 487 GHz transition has  a much larger beam and marginal signal to noise ratio.  

When processed through HIPE version 5, the pointing offsets of the polarizations relative to the nominal pointing direction are recorded.  For March, the $\alpha$ and $\delta$ offsets for H and V are (--0.9\arcsec,+2.1\arcsec) and (+0.9\arcsec,--2.1\arcsec), while for September they are (+0.5\arcsec,--2.2\arcsec) and (-0.5\arcsec,+2.2\arcsec), respectively.  Thus, the offsets from separate polarizations have almost been exactly interchanged due to the 6 months that have elapsed between the two sets of observations.  

The V(March) and H(Sept) pointing direction is in a direction offset towards the Hot Core  by 2.3\arcsec, and the V(Sept) and H(March) data are offset in the opposite direction by an equal amount.  We can investigate whether the measured intensities are consistent with a source for the \oxy\ emission located at the position of the Hot Core.  To do this, it is thus reasonable to average together the H data from September with the V data from March, and the V data from September with the H data from March.  Doing this we find $\langle \int T$[H(Sept) + V(Mar)]$dv \rangle$ = 0.160 $\pm$ 0.023 \kks and $\langle \int T$[V(Sept) + H(Mar)]$dv \rangle$ = 0.210 $\pm$ 0.012 \kks. The ratio of the integrated intensities is $\langle \int T$[V(Sept),~H(Mar)]$dv \rangle$/$\langle \int T$[H(Sept),~V(Mar)]$dv \rangle$ = 1.31$\pm$ 0.20.  

The Hot Core is separated from the \hh\ Peak 1 nominal pointing position by 24.6\arcsec.  Adding the offsets for the different epochs appropriately, we find that the distance for the H(Sept) and V(Mar) data is 22.4\arcsec, and for the V(Sept) and H(Mar) data is 26.9\arcsec.  Taking the Hot Core Source to be small compared to the antenna beam size, the ratio of the signal measured when pointing closer to the source to that found when pointing farther from the source is given by $\exp((d_{farther}/\theta_0)^2 - (d_{closer}/\theta_0)^2)$, where $\theta_0$ is the 1/e radius of the beam, equal to 16.6\arcsec\ at 774 GHz.  The resulting ratio is 2.27.

Relevant information about pointing and source location is provided by the other lines observed simultaneously with the \oxy.  We see changes in most lines, and the \csev\ line is particularly useful as it is relatively isolated as well as quite strong.  It seems reasonable that the broad component of this line originates in a relatively small region near the Hot Core. Table \ref{C17O_wide} includes the data for the two polarizations at two epochs.  The strength of the broad component is sufficient that we find ratios of the integrated intensities (as defined above) V(Mar)/H(Mar) = 1.41 and H(Sept)/V(Sept) = 1.77, with an average value equal to 1.59.  Both ratios are in the sense expected for a point source in the vicinity of the Hot Core, and  ``flip'' in the manner that is expected.  

\begin{deluxetable}{ccc}

\tablecolumns{3}
\small
\tablewidth{0pt}

\tablecaption{\label{C17O_wide} Parameters of Broad Component  of  \csev\ Line For Different Polarizations at Different Epochs}
\tablehead{  \colhead{Epoch}   & \colhead {$\int T_{mb}dv$}         &\colhead{$\sigma[\int T_{mb}dv]$}\\
                    \colhead{}             & \colhead {(K km s$^{-1}$)}          &\colhead{(K km s$^{-1}$)}     
                 }
\startdata
H(Mar)   &  \,\,\,9.54  &0.15\\
V(Mar)   &13.47  &0.37\\
H(Sept)  &22.17  &0.18\\
V(Sept)  &12.51  &0.13\\
\enddata

\end{deluxetable}

The formal statistical error for each ratio above is less than 4 percent, so these two measurements are marginally consistent with their average value.   However, the data in Table \ref{C17O_wide} suggest that there is another systematic difference, which can be seen in the average of H and V in September being significantly greater than the average of H and V in March by a factor of 1.5.  Since the relative pointing offsets of the two polarizations are essentially just interchanged between the two observing epochs, this suggests that there was an overall pointing change between the periods in which the observations were carried out.  

Assuming we are observing a pointlike source at the nominal distance of the Hot Core, we can solve for the pointing offset, which is 2.4\arcsec\ in the sense that the average pointing direction in September was this amount closer to the position of the Hot Core than in March.  Although this error results in a large difference in the signal received when observing a source on the ``side'' of the beam, this is a relatively small pointing error ($\simeq$ FWHM/12), which is within the expected pointing precision for {\it Herschel} \citep{pilbratt2010}.  Our data are only marginally consistent with the \oxy\ emission being a small source at the location of the Hot Core.  The NH$_3$ emission from the northwest portion of the Hot Core, and the Peak A - Western Clump, and the Murata et al.\ CS clumps are all somewhat closer, and thus more consistent with our  measurements.  

\section{Analysis}
\label{analysis}

\subsection{ Collisional Excitation of \oxy}
\label{excitation}
Since the millimeter- and submillimeter-wavelength transitions of \oxy\ are magnetic dipole transitions, their spontaneous decay rates are relatively small.  Thus, in the relatively dense regions being observed, it is likely that collisions are the dominant excitation mechanism.  \citet{orlikowski1985} carried out calculations for collisions between \oxy\ and He atoms (as surrogates for \hh\ in J = 0 level) for levels defined by rotational quantum numbers N = 1 and 3, at a total energy of 300 cm$^{-1}$ (corresponding to 432 K).  \citet{corey1986} performed similar calculations for excitation from N = 1 and 3 to N = 3, 5, and 7 levels for a collision energy of 313 K.  The results suggest a propensity for collisions in which the quantum number F, which is equal to N -- J (and hence = --1, 0, or +1)  does not change.  \citet{bergman1995} and \citet{valkenberg1995} carried out close-coupled calculations of \oxy\ by He over a range of temperatures relevant for interstellar clouds, using a semi-empirical interaction potential and a number of approximations.  

The uncertainty due to the potential energy surface (PES) has been reduced by use of an improved surface in the recent close-coupled calculations by \citet{lique2010}.  The recent study finds that the collisions in which F is unchanged have rate coefficients $\simeq$10$^{-10}$ cm$^3$ s$^{-1}$,  essentially independent of N.  The collisions in which N changes are  a factor of 10 smaller for low N, and almost a factor of 100 smaller for N $\geq$ 10. The transitions with the largest cross sections do not correspond to ones for which radiative transitions are allowed.  Thus, the upper level of an observed transition is generally not populated by a collision from the lower level of the transition.  The result is that a critical density defined in terms of the spontaneous decay rate divided by the deexcitation rate coefficient (for the same levels) is overestimated.

Given that we expect the transitions observed to be optically thin, the most meaningful quantity is simply the fractional population of the upper level, since the observed antenna temperature is directly proportional to this quantity.  Using the rate coefficients  at a kinetic temperature of 100 K from  \citet{lique2010}, scaled by a factor of 1.37 to account approximately for \hh\ rather than He as a collision partner, we find that the upper level fractional populations are within 10\% of their LTE values for densities above 1$\times$10$^3$ \cc\ for the 487 GHz transition, 1.3$\times$10$^4$ \cc\ for the 774 GHz transition, and 2$\times$10$^4$ \cc\ for the 1121 GHz transition.   This density for the 1121 GHz transition is a factor of 6 smaller than the critical density defined in the usual manner.

In moderate density interstellar clouds, the molecular hydrogen density is n(\hh) $\simeq$ 10$^3$ \cc, which is sufficient to bring only the population of the (3,3) level close to LTE; the populations of the upper levels of higher-frequency transitions will be subthermal.  However, in regions such as the centers of giant molecular clouds, where densities are generally $\geq$ 10$^5$ \cc\ and certainly in excess of 10$^4$ \cc, the fractional populations of the upper levels of all transitions will be close to LTE values. 
This is confirmed by a fully non-LTE treatment of the O$_2$ excitation, which produces the same result for column density and temperature when the hydrogen density is $\geq\ 10^4$ \cc.   It is thus appropriate to use the relative line intensities with the assumption of LTE to determine the kinetic temperature and the total \oxy\ column density.

\subsection{\label{temperature} Temperature of the \oxy-emitting Region}

\begin{figure}
\begin{center}
\includegraphics[width = 9cm]{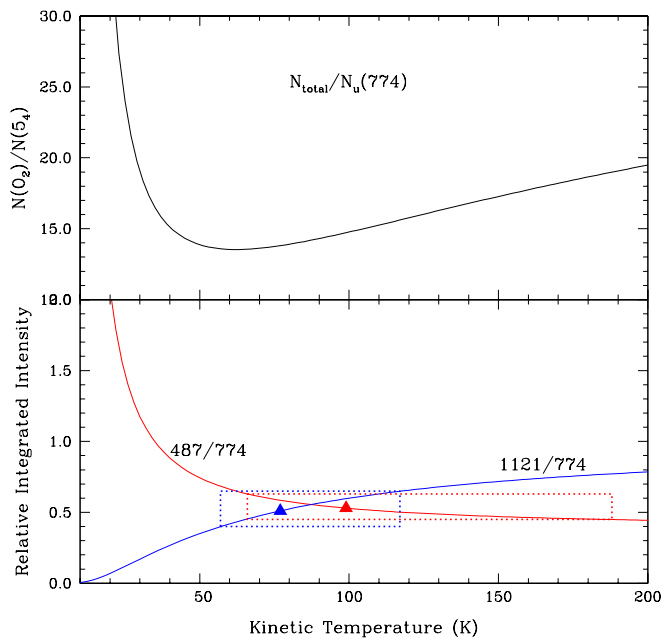}
\caption{Lower panel:  Intensity of the 487 GHz \oxy\ transition relative to the 774 GHz transition, and of the 1121 GHz transition relative to the 774 GHz transition.  These values assume that the beam filling factor is the same for the three transitions, and that all levels are populated according to LTE.  The triangles are from the results given in Table \ref{oxy_param}.  The dotted boxes are defined by the statistical uncertainties in the observed intensity ratios, and the corresponding values of kinetic temperature that are implied.  Upper panel: Total column density of \oxy\ relative to that in the upper level of the 774 GHz transition, (N,J) = (5,4).  In LTE, this depends only on the kinetic temperature as shown here.}
\label{relintens}
\end{center}
\end{figure}

We can use the three transitions observed to determine the temperature of the \oxy-emitting region.  To do this we first assume that the source can be characterized by a single temperature, and that it fills the beam of all three transitions observed.  Both of these assumptions may only hold approximately.  If there is a fractional beam filling factor less than unity, but which is the same for all beams, it does not affect their relative intensities or the beam-averaged column density that we determine.  Assuming  that the source does fill the main beam of all three transitions and that the emission is optically thin, the integrated main beam  brightness temperature is proportional to the upper level column density of the transition observed, $N_u$, through the relationship 

\begin{equation}
\label{coldens}
N_u~(\rm cm^{-2}) = 1.94\times 10^3~\frac{\nu ^2~(GHz)}{A_{ul}~(s^{-1})}  \int T_{\rm mb}dv~(K~km s^{-1})~, 
\end{equation}
where  $\nu$ is the frequency of the transition and $A_{ul}$ is its spontaneous decay rate.  Assuming LTE, as discussed above, the ratio of the column densities of two transitions is determined only by the kinetic temperature of the gas.  With three transitions, we can form two such ratios, and from each determine the range of kinetic temperatures consistent with the observations (including at least statistical uncertainties in $\int T_{mb}dv$).  The results are shown in Figure \ref{relintens}.  The data in Table \ref{oxy_param} yield 0.45 $\leq$ I(487)/I(774) $\leq$ 0.63 and 0.40 $\leq$ I(1121)/I(774) $\leq$ 0.65.  These limits define the temperature limits shown in Figure \ref{relintens}. A kinetic temperature between 80 K and 100 K seems most likely, but values between 65 K and 120 K are consistent with the relative intensities of all three transitions observed.  

From equation \ref{coldens}, we  obtain that N(5,4) = 4.5 $\pm$ 0.21 $\times$ 10$^{15}$ cm$^{-2}$.  With the kinetic temperature defined consistently by the two ratios of the three integrated intensities, the total \oxy\ column density relative to that in (N, J) = (5,4) can be reliably calculated, and the result is shown in Figure \ref{relintens}.  The energy of the (N, J) = 5, 4 level is such that there is remarkably little variation in the fraction of the total \oxy\ in this one level.  We combine the uncertainty in N(5,4) with that in the kinetic temperature to determine the beam-averaged column density N(\oxy) = 6.5 $\pm$ 1.0 $\times$10$^{16}$ cm$^{-2}$.

\subsection{Filled Beam Fractional Abundance of \oxy}
\label{xo2} 
Due to the wide range of temperatures and kinematic variations in the gas towards \hh\ Peak 1, it is difficult to determine the \hh\ column density to compare with that of the \oxy\ observed.  The \csev\ J = 7 -- 6 transition discussed in \S\ref{velloc} is a good candidate for this, since it is optically thin, and at these elevated temperatures, depletion of carbon monoxide should not be significant.  The width of this line is sufficient to include all hyperfine components, but as discussed in \S\ref{velloc}, we do see three velocity components in addition to the very wide component, presumably associated with the molecular outflow.  A four-Gaussian fit is very satisfactory, and yields the results given in Table \ref{C17O_7-6_param}.  While the velocity of component 4 (10.3 \kms) agrees best with that of the \oxy, the line width (1.3 \kms) is considerably narrower than the $\simeq$ 3 \kms\ measured for the three \oxy\ lines (Table \ref{oxy_param}), so it is difficult to make a definitive correspondence.  

Using equation \ref{coldens}, we find J = 7 \csev\ column densities 3.3$\times$10$^{13}$ cm$^{-2}$ for the 10.3 \kms\ component and 7.8$\times$10$^{14}$ cm$^{-2}$ for the combination of the three narrow components (2 to 4).  If we consider the emission at velocities above 10 \kms\ excluding the broad components 1 and 2, we obtain a column density equal to 2.5 $\times$10$^{14}$ cm$^{-2}$.  
The fraction of \csev\ in J = 7 becomes almost independent of temperature for T $\geq$ 80 K(varying from 1/13 at 80 K to a minimum of 1/10 at 150 K, and increasing very slightly higher at temperatures below 300 K).  We adopt a value of 1/11, which yields N(\csev) summed over all levels equal to 3.6$\times$10$^{14}$ cm$^{-2}$ for the very narrow component and 8.7$\times$10$^{15}$ cm$^{-2}$ for the narrow components combined.  We take the ratio CO/\csev\ = 2460 \citep{penzias1981}, and \hh/CO = 10$^4$, yielding N(\hh) = 8.9$\times$10$^{21}$ cm$^{-2}$, 6.7$\times$10$^{22}$  cm$^{-2}$ and 2.1$\times$10$^{23}$ cm$^{-2}$, for the 10.3 \kms\ feature, the $\geq$10 \kms\ emission, and for the combined narrow components, respectively.  

This range of column densities corresponds to a range of densities averaged over the projected diameter of the largest {\it Herschel} beam (44\arcsec) of
$ \langle n({\rm H}_2)\rangle = N({\rm H}_2) / \theta D = 3\times 10^4\;{\rm to}\;
8\times 10^5\;\; {\rm cm}^{-3}. $
The distance $D=416$ pc is the weighted mean of four recent parallax measurements of masers in the Orion KL region and of stars in the Orion Nebula Cluster \citep{hirota2007, sandstrom2007, menten2007, kim2008}.  These densities are in harmony with the requirements of the preceding LTE analysis of the O$_2$ line intensities.  With these values bracketing the range of \hh\ column density associated with the \oxy\ emission, we find that the fractional abundance of \oxy\ is between 3.1$\times$10$^{-7}$ and 7.3$\times$10$^{-6}$.   The higher value is still only $\sim$ 1\% of the total oxygen available.  The lower value is comparable to the upper limits found by {\it SWAS} and {\it Odin}.  if the fractional abundance of carbon monoxide is lower (which is plausible, the adopted value being close to the maximum observed), the column density of \hh\ will be increased and the fractional abundance of \oxy\ proportionately reduced and thus be even smaller compared to the total elemental oxygen abundance.

\section{Beam Filling and Source Offset Effects}
\label{filling_offset_effects}
Modeling of the emission for anything but an extended source with a single filling factor for all transitions involves source coupling to beams of the three transitions each having a different beam size.  This leads to some effects which we discuss here.

\subsection{Effect of Beam Filling}
\label{beam-filling}
A beam filling factor less than unity, but which is the same for all transitions, can be produced by an extended but clumpy medium (as suggested by e.g. the shocked \hh\ emission seen in Figure \ref{orion_hifi_beams}).  This situation leaves the relative intensities of the transitions unchanged, and thus the temperature we determine for the region responsible for the \oxy\ emission will be unaffected.  The column density of \oxy\ in the ``filled'' portion of the beam is increased by the reciprocal of the beam filling factor.  If the \oxy\ emission is associated with postshock gas traced by \hh\  emission, then as shown in Figure \ref{orion_hifi_beams}, the emitting region is extended on the scale of all of the HIFI beams, albeit modestly weaker in the outer portions of the lowest frequency (487 GHz) beam.  The fractional abundance of \oxy\ may be increased if the tracer of \hh\ is uniformly distributed, but may be essentially that derived above if the overall mass distribution is nonuniform, rather than only the distribution of \oxy.  

If Peak A is the source of the observed \oxy\ emission, we have to deal with a small source which is also away from the antenna boresight direction.  In Appendix 2 we discuss the limiting case of a source that is so small that it can be considered to be point-like relative to all beams.  The ratio of the intensities of two transitions, 2 and 1, can be considered to be the ratio characteristic of the source itself multiplied by the correction factor $R(2/1)$ as defined in equation \ref{small_21} in Appendix 2.  To derive the required characteristics of the source, we must thus divide the observed ratio by the correction factor.  For the ratio of the 1121 GHz transition to the 774 GHz transition we find
 
\begin{equation}
\frac{I^{obs}(1121)}{I^{obs}(774)} =  \frac{I^{src}(1121)}{I^{src}(774)} [\frac{\lambda(774)}{ \lambda(1121)}]^2  = 2.1 \frac{I^{src}(1121)}{I^{src}(774)}\lp
\end{equation}
The ratio for the source has to be a factor $\simeq$ 2 smaller than the observed ratio, which as seen in Figure \ref{relintens}, means a lower kinetic temperature.  For an observed ratio of 0.5, a temperature of 70 K is required for a source filling both beams.  However, the ratio for a small source must be smaller by a factor of 2.1, and hence equal to 0.24.  As seen from Figure \ref{relintens}, this requires a kinetic temperature of $\simeq$ 35 K.

The second line ratio is again normalized to the intensity of the 774 GHz transition, and hence the correction factor $R(487/774)$ = $ [\lambda(774)/\lambda(487)]^2$ = 0.4.   As indicated by the appropriate curve in Figure \ref{relintens}, this again corresponds to a lower kinetic temperature for the small source.  An observed ratio of 0.5, which requires a filled source temperature of 100 K demands a ratio 2.5 times larger in a small source, and thus a source temperature again $\simeq$ 30 K.  A small source on boresight is consistent with a single temperature determined by the ratios of integrated intensities, but the source temperature is too low given the ease of excitation of \oxy\ and the temperature of the center of the Orion molecular cloud.  It is thus unlikely that the observed \oxy\ lines are produced by a small source coincident with Orion \hh\ Peak 1, but an extended while beam-diluted source can  fit the present data.

\subsection{Effect of Source Offset}
\label{offset} 
We again restrict our discussion to a small (point-like) source, which when offset by an angle $\theta$ produces a signal reduced from that of a source on boresight by a factor which is given by the normalized antenna response pattern

\begin{equation}
P_n = e^{-(\theta / \theta_0)^2}  \lc
\end{equation}
where $\theta_0$ (= 0.601$\theta_{FWHM}$) is the 1/e beam radius.  $\theta_0$ is proportional to the observing wavelength, and so a given offset produces a greater reduction in observed integrated intensity at a higher frequency.   Following the discussion of beam filling, we define the  integrated main beam temperatures with the source off and on boresight as $I^{\rm off}$ and $I^{\rm on}$, respectively.   We then find
\begin{equation}
\frac{I^{\rm off} (1121)}{I^{\rm off} (774)}  = \frac{e^{(\theta / \theta_0(774))^2}}{e^{(\theta / \theta_0 (1121))^2}}~ \frac{I^{\rm on} (1121)}{I^{\rm on} (774)} \lp
\end{equation}

Since $\theta_0(1121)$ = 11.4\arcsec\ and $\theta_0(774)$ = 16.6\arcsec\, the correction factor for off-boresight pointing is always less than unity and decreases as the offset increases.   To fit a given observed ratio, we require an on-boresight source with a larger ratio, which can only be produced by a higher source temperature.  The ratio for the 487 GHz transition to the 774 GHz transition shows the opposite dependence, being greater than unity.  Thus, to fit an observed integrated intensity ratio, the off-boresight source would need to have a smaller 487 GHz to 774 GHz ratio than that for an on-axis source, and the effect again is that the source temperature would be greater than if the source were on boresight.  

The observed 1121 to 774 ratio, close to 0.5, is what would be produced by a on-boresight small source that if filling both beams would produce a ratio equal to 0.5/2.1 = 0.24.  To obtain a ratio for a small source off-boresight equal to the on-boresight ratio for a source that fills the beams requires that the off-borseight correction be the inverse of the small-source correction, and thus equal to 1/2.1 = 0.48, which occurs if the source is off axis by just over 13\arcsec.  

As seen in Figure \ref{relintens}, the ratio $I^{src} (1121)/I^{src} (774)$ has an upper limit of about 0.8 for a filled source, irrespective of temperature.  This would be produced by a on-boresight small source with an (unphysical) ratio of 1.68.  However, a \textit{very warm small source off-boresight} could produce the observed ratio of 0.5 with off-boresight correction factor equal to 0.30, which results when the source is off  boresight by about 17\arcsec.  
Especially when the uncertainties are considered, consistent solutions for all three transitions are possible for small sources over a restricted range of offset angles, up to approximately this limit.
Since the offset to the Hot Core is close to 27 \arcsec, it appears difficult to have this source dominate the \oxy\ emission we have observed at \hh\ Peak 1.
One or more of the compact sources discussed previously are more likely the source of the emission. In particular, Peak A (with an offset of 21\arcsec), is consistent with this picture, with gas and dust temperatures well over 100 K being indicated by other observational data and required to fit the present observations of \oxy.  

\section{Discussion}
\label{discussion} 
There seems little reason to doubt that we have detected emission from \oxy\, but the source of the emission remains imperfectly characterized.  The issue at hand is to understand the circumstances that result in the increase in the \oxy\ abundance from the low levels derived from {\it SWAS} and {\it Odin} observations.  Two theories for enhanced abundance of \oxy, (1) warm dust with restoration of pure gas-phase chemistry, and (2) shocks, seem to be relevant possibilities for explaining the present observations. We here give some additional details that bear on explaining our data for \hh\ Peak 1 in Orion.

\subsection{Warm Dust}
\label{dust} 
It is generally accepted that dust grains in dense clouds without star formation in the immediate vicinity will  be sufficiently cold that over time (depending  on the density), atomic oxygen will deplete onto grains and to a large degree be hydrogenated to water.  In consequence, the grains acquire water ice mantles incorporating a significant fraction of the available oxygen, with an accompanying significant reduction in the gas-phase X(\oxy),  as developed in the models by \citet{bergin2000} and \citet{hollenbach2009} discussed in \S\ref{intro}.  The formation of a massive star in or near the region in question produces radiation that is in a short distance degraded to infrared wavelengths, and which heats dust grains.  At modest temperatures, any desorbed \oxy\ will be returned to the gas, as discused in \S \ref{oxy_chem}.  When the grain temperature reaches $\sim$ 100 K, return of water ice to the gas phase becomes rapid \citep{fraser2001}.  This starts the restoration of ``pure'' gas-phase oxygen chemistry.  The time required to reach steady state depends primarily on the ionization rate, but ultimately a very substantial X(\oxy) $\geq$ 10$^{-5}$ will be established.   

\begin{figure}[h!]
\begin{center}
\includegraphics[width = 9cm]{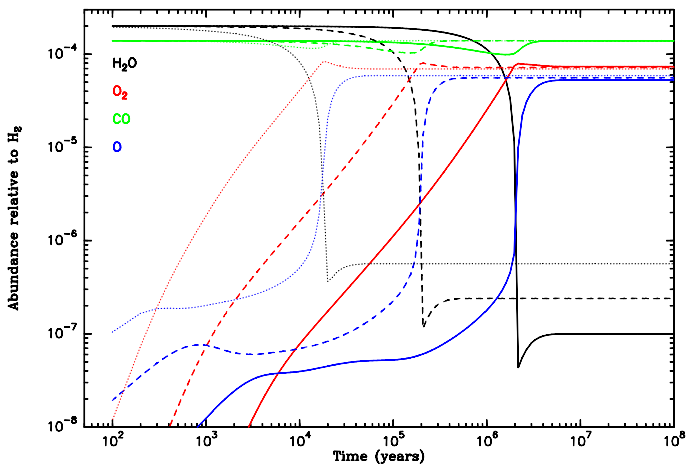}

\caption{\label{oxy_tdep_fig}Time evolution of gas-phase abundances of atomic oxygen, molecular oxygen, water, and carbon monoxide in a cloud with dust and gas temperatures equal to 100 K and  \hh\ density equal to 10$^6$ \cc.  At t = 0, elemental oxygen not in CO is almost entirely in the form of gas-phase water, grain surface water having been returned to the gas phase as a result of the warm grain temperature.  The dotted curves are for a cosmic ray ionization rate of 10$^{-15}$ s$^{-1}$, the dashed curves for 10$^{-16}$ s$^{-1}$, and the solid curves for 10$^{-17}$ s$^{-1}$.  }

\end{center}
\end{figure}

In Figure \ref{oxy_tdep_fig}, we show results of a representative calculation, which is based on NAHOON, the \citet{wakelam2005} model, but using the OSU--1--2009 reaction rate file.   The \hh\ density is 10$^6$ \cc, the gas temperature is 100 K, and grain surface reactions are excluded except for \hh\ formation.  We assume that at t = 0, all of the elemental oxygen that had been on the grain surface is returned to the gas phase, and that the oxygen that had (in the steady-state solution for large times in this model) been in molecular and atomic form, as well as that having been water ice, comes off the grain as H$_2$O.  This reflects the idea that at some point in the past, the grains had been cool enough that all of these species would deplete onto the grain surfaces, and end up as water ice, but the result here is in effect a gas-phase chemistry model.  There are also trace oxygen-bearing species such as SO and SO$_2$ that are in the gas phase at the start of the present calculation, but together they represent less than 0.5\% of the total available oxygen.

Starting at t = 0, the \oxy\ fractional abundance builds up steadily, and although initially smaller than that of atomic oxygen, soon surpasses it.  The two species end up with comparable steady-state abundances 5 to 7 $\times$10$^{-5}$.   The time to achieve maximum abundance of \oxy\  scales inversely with the cosmic ray ionization rate.  The time scales for achieving a given \oxy\ fractional abundance as well as steady-state abundance are largely independent of the density of the region.  The behavior is also largely independent of temperature as long as the dust is warm enough to prevent any significant grain surface depletion to occur.  The steady-state fractional abundance of H$_2$O increases as the ionization rate increases, although the abundances of O and \oxy\ are only slightly affected.

To evaluate the quantity of gas that may be maintained at conditions for the above chemistry to evolve, we evaluate a highly idealized model based on a single luminous source in a spherically symmetric cloud.  Taking a central source with 10$^4$ -- 10$^5$ L$_{\sun}$, plausibly IRc2 \citep{downes1981} or else a collection of sources in the region \citep{gezari1998} having similar total luminosity, we have carried out a representative calculation of the dust temperature distribution using the DUSTY radiative transfer code (http://www.pa.uky.edu/$\sim$moshe/dusty;  Nenkova et al.\ 2000).   The center to edge visual extinction is equal to 1000 mag., but this value is unimportant as long as the dust at the cloud edge is cooler than the temperatures of interest.  We find that the solutions at distances of interest are quite independent of the radial density distribution assumed to be a power law $n(r)$ $\sim$ $r^{-\alpha}$ with index $\alpha$ between 0 and 2.  At the 4$\times$10$^{16}$ cm projected distance of Peak A from IRc2, the dust temperature is 56 K for the lower luminosity and 100 K for the higher.  At the 1.5$\times$10$^{17}$ cm projected distance of \hh\ Peak 1 from IRc2, the temperatures are 35 K and 55 K, respectively.  Thus, even if Peak A is externally heated, the grains will be sufficiently heated to result in ``warm dust'' chemistry, while those at the distance of \hh\ Peak 1 will not.

The fraction of the total opacity between the central source and the location of a particular grain temperature does depend on $\alpha$.  For relatively flat density distributions ($\alpha$ $<$ 1), the fraction of the total optical depth is $\leq$ 0.5 for a dust grain temperature of 100 K.  While recognizing that this idealized, spherically symmetric model may not be accurate for this complex region, it seems plausible that IRc2 or another source, or the combination of sources in the region can result in warm grains in Peak A, even if this source is not self-luminous.  The greater distance to \hh\ Peak 1 means that the grains there would still be cold enough to retain significant ice mantles if IR heating were the only operative process warming the dust.

Given that the grains in Peak A may be sufficiently warm that the gas-phase \oxy\ abundance has been significantly raised, we can examine whether this source is consistent with the \oxy\ line intensities we have observed.  We must thus consider a small source offset from the beam pointing direction.  Following the discussion in \S\ref{beam-filling} and \S\ref{offset}, the 1121/774 ratio is increased by a factor of 2.1 due to small source size and decreased by factor 0.15 due to offset, relative to filled-beam on-boresight model.  The resulting correction factor is 0.3, which means that the maximum ratio for a warm beam-filling source of 0.8 becomes 0.25, lower than the observed ratio of 0.51$\pm 0.1$.  For the 487/774 ratio, we get a combined correction factor of 1.12, which when multiplied by the high temperature filled and centered model ratio of 0.45, gives a ratio equal to 0.51, compared to the observed ratio of 0.54$\pm$0.1.  Considering the uncertainties in source size and geometry, in addition to calibration issues, we conclude that the observed line ratios are consistent with a small warm source located at the position of Peak A.  

If Peak A is the source of the 11 \kms\ to 12 \kms\ emission, the column densities are certainly greater than the filled beam values given previously. With a representative 10\arcsec\ for the  diameter of Peak A (larger than \citet{murata1992}, as there is some extended emission) and including the attenuation due to being 21\arcsec\ away from boresight, we obtain a total correction factor of 55 compared to a filled-beam on-boresight source.  The \csev\ used to determine the associated \hh\ column density certainly does not all come from this small source, but scaling  the \hh\ column density implied by the $\geq$ 10 \kms\ emission, 6.7$\times$10$^{22}$ cm$^{-2}$, by the correction factor, we find N(\hh) = 3.7$\times$10$^{24}$ cm$^{-2}$.  The conversion from \csev\ in J = 7 to total N(\csev) is essentially independent of temperature for T $\geq$ 100 K.  The origin of the \oxy\ emission in this small region leaves the emission optically thin, as is the \csev\ emission for v $\geq$ 10 \kms.  For a source temperature $\geq$ 150 K compared to the $\sim$ 100 K value in the filled-beam model, the column density of \oxy\ will be slightly increased due to the larger partition function  (Figure \ref{relintens}).  For a temperature of 180 K we obtain N(\oxy) = 4.6$\times$10$^{18}$ cm$^{-2}$ for a 10\arcsec\ diameter source.  The fractional abundance of \oxy\ is only slightly changed and is equal to 1.3$\times$10$^{-6}$, a value consistent with the warm grain chemical models.  According to the warm dust model as shown in Figure \ref{oxy_tdep_fig}, for a low ionization rate of 10$^{-17}$ s$^{-1}$, the time to reach this fractional abundance is $\leq$ 10$^5$ yr.  

\citet{murata1992} derive a mass of 3.2 M$_{\sun}$ for Cnt D clump, coincident with the  Peak A -- Western Clump -- MF4 source, as discussed above.  \citet{wright1992} carried out submillimeter dust observations with somewhat lower angular resolution, and derived a size of 5\arcsec\ -- 10 \arcsec\ and a mass of  4 -- 13 M$_{\sun}$, slightly larger than those of \citet{murata1992}, possibly due to their lower angular resolution.  The peak column density from the \citet{murata1992} data is N(\hh) = 5$\times$10$^{24}$ cm$^{-2}$, which agrees surprisingly well with that implied by our \csev\ and related observations keeping in mind that the mass derived from submillimeter dust emission has its own $\sim$ factor of three uncertainty.  The average density in the clump is $\simeq$ 2$\times$10$^{8}$ \cc, which is certainly adequate to thermalize both the \oxy\ and the \csev.  

The simplest estimate for the virial mass of this clump, assuming uniform density, and ignoring magnetic fields and external pressure, is M$_{vir}$ = 5$\sigma_{los}^2$R/G, where $\sigma_{los}$ is the line of sight velocity dispersion and R is the cloud radius.  The result is M$_{vir}$ = 8 -- 19 M$_{\sun}$, and M$_{vir}$/M = 0.7 -- 4 for the range of sizes and masses determined from the dust emission.  This range includes values characteristic of cores in general.   While we cannot definitively prove that the observed \oxy\ lines arise from this compact source, it certainly does seem to be a good candidate to explain our observations.

\subsection{Shocks}
\label{shocks} 

The \hh\ Peak 1 position was selected as it coincides with the strong local maximum in the intensity of the v = 1 -- 0 S(1) transition of molecular hydrogen studied by \citet{beckwith1978}.  Using several of the vibration--rotation transitions, these authors found an excitation temperature in excess of 2000 K.   Since the peak is not in the immediate vicinity of any luminous source, this very high value is a strong indication that this is a result of shock excitation.  
In the intervening years there has been a plethora of papers that report dramatically improved spatial and spectral resolution \citep{scoville1982, chrysostomou1997, schild1997, stolovy1998, schultz1999, salas1999, rosenthal2000, vannier2001, gustafsson2003, kristensen2003, lacombe2004, bally2011}.

Although fluorescence powered by UV radiation from the Trapezium stars dominates the H$_2$ line emission on degree scales in Orion \citep{luhman1994}, the emission of highest surface brightness in the molecular core is due mainly to collisional excitation in shocks \citep{rosenthal2000}.  As discussed by these authors, population diagrams based on the wide range of lines available from ISO clearly indicate a large range of excitation temperatures spanning a few hundred to several thousand K.
Multiple shocks are indicated by the pure rotation \hh\ transitions observed by \citet{parmar1994}.  A relatively low-temperature component is also indicated by the 17 \mic\  J = 3 -- 1 rotational transition studied by \citet{burton1997}.  This study also suggested the presence of shocks having velocities as low as 5 \kms.  

High resolution observations provide evidence of a complex kinematic structure.  Among the highest spectral resolution data are those of \citet{chrysostomou1997} of the \hh\ v = 1 -- 0 S(1) line, carried out with a velocity resolution of 14 \kms. These data indicate that the spatially extended \hh\  emission peaks at a velocity of 12 \kms\ and has an average FWHM line width equal to 37 \kms, but that localized regions have different central velocities and broader line widths.  The highest spatial resolution observations \citep[0.15\arcsec,][]{gustafsson2003} show complex localized structures having dramatically different central velocities.

The effects of shocks on the abundance of \oxy\ have been considered in some detail by \citet{bergin1998}; references to reaction rates and earlier studies can be found in that paper.  The key point is that the reaction O + \hh\ $\rightarrow$ OH + H, which has an activation barrier of 3160 K, can become the dominant production pathway for OH at postshock gas temperatures.  This provides a major additional source of OH beyond the primary pathway operative in quiescent clouds in which dissociative recombination of H$_3$O$^+$ is the major source of OH.   Thus, if there is appreciable preshock atomic oxygen, there will be enhanced production of OH in the shock, which then can react with O via the temperature-independent reaction OH + O $\rightarrow$ \oxy\ + H discussed in \S\ref{intro}.  FIR observations \citep{goicoechea2006} suggest that there is large increase in the abundance of OH as a result of the shock in the Orion KL outflow, but the velocity and line width of the OH do not allow any conclusive association with the \oxy\ emission.

However, the reaction OH + \hh\ $\rightarrow$ H$_2$O + H, with an activation barrier of 1660 K, competes for the available OH.  In addition, there is a back reaction H + O$_2$ $\rightarrow$ OH + O, which limits the abundance of \oxy\ if the gas temperature exceeds $\simeq$ 1000K and there is some atomic hydrogen present.  These conditions are likely to occur following the passage of a strong nondissociative shock.  Thus, shocks strong enough to enable production of OH, but in which the postshock temperature does not exceed $\simeq$ 500 K, result in maximum \oxy\ production \citep{bergin1998}.  \citet{kaufman2010} has studied a wide range of nondissociative C--shocks over a wide variety of shock velocities  and preshock densities, and finds that maximum \oxy\ production results from shock velocities 10 \kms\ to 12 \kms, and is relatively independent of preshock gas density.  The range of shock velocities over which \oxy\ column densities in excess of 10$^{16}$ cm$^{-2}$ are produced is broadened for n(\hh) in the preshock gas $\ge$ 10$^4$ \cc.  Under optimal conditions, \oxy\ column densities can be as large as 10$^{17}$ cm$^{-2}$. 
 
The models of shock chemistry, while still highly idealized relative to the complex structure in the \hh\ Peak 1 region (see e.g. Figure \ref{orion_hifi_beams}), may well produce a relatively large column density of \oxy.  The conditions required are (\textit{i}) a C--shock of appropriate velocity, (\textit{ii}) low electron density with X(e$^-$) $\leq$ 10$^{-6}$ -- 10$^{-7}$, (\textit{iii}) sufficiently high preshock density, n(\hh) $\geq$ 10$^4$ \cc, and (\textit{iv}) a significant fraction of available oxygen in gas-phase atomic form.  The first three requirements do not seem implausible in this region of Orion, due to the high degree of clumpiness, the  high average density,  and the wide range of shock velocities present.   In considering shock models, we have to allow for a range of shock velocities propagating in a very nonuniform medium.  This is to some degree the picture advanced by \citet{salas1999} in which shocks driven by a stellar wind impinge on (possibly moving) clumps.  The shock velocity in a dense clump will be significantly lower than that in the diffuse interclump medium, and so one can easily imagine a wide range of shock velocities.  In addition to the shock propagating into the clump (which may have too low a velocity to produce abundant \oxy), there can be a variety of other complex shocks that could release atoms and molecules from the grains, which could subsequently form large amounts of \oxy.

The issue of the atomic oxygen abundance is perhaps the most challenging one.  For C--shock speeds in excess of $\sim$ 15 \kms, the ion--neutral streaming is sufficient to sputter grain mantles and thus produce a large density of atomic oxygen.  Even in the absence of sputtering, if  the extinction of this region relative for the UV field produced by the Trapezium stars is not greater than $\sim$ 9 mag, the atomic oxygen fractional abundance could exceed 10$^{-5}$, which would be sufficient for the C--shock to produce observed column density of \oxy.   The \oxy\ column density in the PDR itself ($\leq$ 10$^{16}$ \cc; Hollenbach et al. 2009) is significantly below that observed unless there is some very special geometry involved.   Alternatively, a higher-velocity J--shock (such as suggested by the \hh\ data) could produce the atomic oxygen.   In the dense postshock gas dissociated \hh\ reforms quickly.  A lower-velocity C--shock propagating into this gas would then produce the \oxy\ as discussed above.  It is also possible that UV radiation generated in the J--shock can, through photo--reactions induced in water ice molecules on grains, result in $\simeq$ 10\% of the molecules desorbed being \oxy\ (\citet{oberg2009a} and \"{O}berg, private communication 2011), which could be a supplement to the other production channels.  

The line profile of the \oxy\ produced in a shock depends in detail on the geometry, but for the idealized case of a planar low velocity shock, has a quite narrow line ``core''  together with a broad ``wing''  profile.  In order for the line velocity to be shifted from that of the ambient material in the center of Orion by only 3 to 5 \kms, the shock direction would have to be close to the plane of the sky.  However, if an initially planar shock encounters inhomogeneities, the propagation direction will be modified, and it seems inevitable that some line broadening would result.   The observed 11 -- 12 \kms\ velocity is shifted relative to the 8 -- 9 \kms\ velocity of the ambient gas in the opposite sense of the CO outflow, which has its blue-shifted wing extending towards \hh\ Peak 1 \citep{erickson1982}.  

A serious problem with the shock model is that while the observed N(\oxy), 6$\times$10$^{16}$ cm$^{-2}$, is $\simeq$ a factor of two less than the maximum N(\oxy) that can be produced, the observed column density is a  beam-averaged value.  Thus, we require the shock to essentially fill a $\simeq$ 30\arcsec\ region, corresponding to $\simeq$ 2$\times$10$^{17}$ cm in size at the 416 pc distance of Orion.  Given the highly inhomogeneous structure of the \hh\ emission seen  in Figure \ref{orion_hifi_beams}, it is difficult to see how our three HIFI beams could be filled with gas that has been through a low-velocity shock all giving projected velocities within a $\pm$1.5 \kms\ range.  It is possible that the observed emission could be the result of shocks having passed through beam-diluted clumps distributed throughout the beam, thus allowing a reasonable agreement with the relative intensities we have observed.  The \oxy\ column density would, however, have to be higher by the inverse of the beam-filling factor. If this factor were as large as few, it would put N(\oxy) close to the upper limit for shock production of this species.

In summary, while shock models can in principle produce the observed column density of \oxy, there are difficulties regarding the initial conditions, particularly the abundance of atomic oxygen, and the observed narrow line profiles.  These can be addressed only by more detailed and extensive modeling.

\section{Conclusions}
\label{conclusions} 

We have observed three submillimeter rotational transitions of the \oxy\ molecule in emission, using the HIFI instrument on the {\it Herschel Space Observatory}.  The pointing position was towards Peak 1 of the \hh\ emission in Orion.  Signals were detected from the 487 GHz, 774 GHz, and 1121 GHz transitions, having peak velocities between 11 \kms\ and 12 \kms, and FWHM line widths $\simeq$ 3 \kms.  The detection of three transitions and the elimination of the most obvious possible ``interloping'' lines confirm that the emission is from \oxy.  The determination of the source of the emission and its characteristics are difficult due to the complexity of the region, but the highlights of the analysis are the following.

\begin{enumerate}

\item  Assuming that the emission fills the beam at all three frequencies, the relative intensity of the \oxy\ lines, taken to be optically thin and produced in LTE, suggests that the temperature of the region is between 65 K and 120 K, and that the column density is N(\oxy) = 6.5$\pm$1$\times$10$^{16}$ cm$^{-2}$. Observations of \csev\ J = 7 -- 6 line obtained simultaneously with those of \oxy\ indicate emission in the same velocity range as observed for \oxy. Based on the \csev\ emission at v $\geq$ 10 \kms\ and assuming X(CO) = 1$\times$10$^{-4}$, we derive a \hh\ column density equal to 6.7$\times$10$^{22}$ cm$^{-2}$ and a fractional abundance X(\oxy) $\simeq$ 10$^{-6}$.

\item  The line centroid velocity is close to that of a number of species in a condensation $\simeq$ 6\arcsec\ west of IRc2, called Peak A, the Western Clump, or MF4 by various authors.  Molecules including HDO, HC$_3$N, NO, and HCOOCH$_3$ have peak emission at velocities between 10 \kms\ and 11.4 \kms\ towards this condensation, very similar to the observed velocities of the \oxy\ lines.  This is also the Cnt D localized maximum in the dust continuum emission.

\item  With only one pointing direction, we do not have definitive information on the exact source of the emission.  The position of Peak A is $\simeq$ 21\arcsec\ from the direction of the present observations.  The differential pointing of the two HIFI polarizations suggests that Peak A is more likely the source of the emission than the Hot Core, which is farther away from the beam pointing direction.     

\item  If the  5\arcsec\ to 10\arcsec- diameter Peak A is the source of the observed \oxy\ emission, the relative line intensities require that the source be quite warm, with T $\geq$ 180 K, and the source column density N(\oxy) = 4.6$\times$10$^{18}$ cm$^{-2}$.  If the aforementioned \csev\ emission also originates in this clump, our \csev\ data imply an \hh\  column density equal to 3.7$\times$10$^{24}$ cm$^{-2}$.  This is very similar to the \hh\ column density derived from interferometric observations of the dust in this clump, N(\hh) = 5$\times$10$^{24}$ cm$^{-2}$.  The resulting fractional abundance ix X(\oxy) = 1.3$\times$10$^{-6}$, which is readily obtained in pure gas-phase chemistry.

\item  Explanations for the relatively large \oxy\ fractional abundance compared to the results found by \textit{SWAS} and \textit{Odin} satellites include (\textit{i}) increase in gas-phase \oxy\ as a result of warm grains having desorbed water ice mantles with subsequent reestablishment of standard gas-phase chemistry and (\textit{ii}) enhancement of X(\oxy) by shocks.  

(\textit{i})  Heating dust grains to the $\sim$ 100 K temperature required to desorb water mantles may have resulted from radiative heating, plausibly by nearby luminous sources.  In this case, the observed \oxy\ lines could be emission from Peak A (possibly with contributions from other condensations that are all close enough to IRc2, IRc7, and other heating sources to have T$_{grain}$ $\geq$ 100 K).  The fractional abundance of \oxy\ for this model implied by our observations is 1.3$\times$10$^{-6}$, which is more than an order of magnitude below that achievable.  The time required to achieve the observed fractional abundance of \oxy\ is moderate, $\leq$ few $\times$ 10$^5$ yr even for low cosmic ray ionization rate and varying inversely with this rate.  Since desorption of water mantles is rapid, the total time for even the first high-mass star to produce a fractional abundance consistent with that observed, is compatible with the lifetime of high-mass stars. 

(\textit{ii}) A 10 \kms\ to 12 \kms\ shock can produce the observed \oxy\ column density only if the emission is not significantly beam diluted.  If the shock propagation direction is largely in the plane of the sky, the velocity centroid and line width may be consistent with what is observed.  Shocks are generally assumed to be responsible for the widespread \hh\ emission observed in this region, but generally taken to have much higher velocities.  J--shocks may produce  UV radiation that maintains a significant abundance of atomic oxygen. This is required to be present in the gas which is then subjected to lower-velocity shocks which are optimum for producing the observed \oxy\ column density.  Shocks may also explain the water maser emission observed with {\it Herschel}. In this scenario, the \oxy\ emission would have to come from numerous clumps distributed throughout the {\it Herschel} beams, possibly the same condensations responsible for the highly inhomogeneous \hh\ emission revealed by high angular resolution observations.  

The \oxy\ velocities and widths, association with the high-column density warm Peak A source, and the plausibility of the warm dust model suggest that this compact condensation is responsible for the \oxy\  emission we have observed.

\item  The resolution of the origin of enhanced \oxy\ abundance will require both additional modeling and further observations. Our data do not suggest that the \oxy\ abundance is sufficiently large to resolve the ``oxygen problem'', referring to the question of allocating the available oxygen among different forms (grains, ices, O, H$_2$O, CO, and other molecules; Whittet 2010).  Future observations will hopefully yield a more complete understanding of  the astrochemistry of \oxy\ and the distribution of oxygen in dense molecular clouds.

\end{enumerate}

\vspace{1cm}

We are indebted to the many people who worked so hard and for so long to make the {\it Herschel} mission and the HIFI instrument a success.  HIFI has been designed and built by a consortium of institutes and university departments from across Europe, Canada and the US under the leadership of SRON Netherlands Institute for Space Research, Groningen, The Netherlands with major contributions from Germany, France and the US. Consortium members are: Canada: CSA, U.Waterloo; France: CESR, LAB, LERMA, IRAM; Germany: KOSMA, MPIfR, MPS; Ireland, NUI Maynooth; Italy: ASI, IFSI-INAF, Arcetri-INAF; Netherlands: SRON, TUD; Poland: CAMK, CBK; Spain: Observatorio Astron\'omico Nacional (IGN), Centro de Astrobiolog\'{\i}a (CSIC-INTA); Sweden: Chalmers University of
Technology - MC2, RSS \& GARD, Onsala Space Observatory, Swedish National Space Board, Stockholm University - Stockholm Observatory; Switzerland: ETH Z\"urich, FHNW; USA: Caltech, JPL, NHSC.The \oxy\ excitation calculations were carried out using the Radex code \citep{vandertak2007}.  We appreciate the effort that went into making critical spectroscopic data available through the Jet Propulsion Laboratory Molecular Spectroscopy Data Base (http://spec.jpl.nasa.gov/), the Cologne Database for Molecular Spectroscopy, (http://www.astro.uni-koeln.de/cdms/ and \citet{mueller2001}) and the Leiden Atomic and Molecular Database (http://www.strw.leidenuniv.nl/$\sim$moldata/ and \citet{schoier2005}).   We thank Holger M\"{u}ller for helpful discussions about molecular spectroscopy.   Colin Borys of the  NASA Herschel Science Center gave us valuable assistance in unravelling the pointing offsets of the two HIFI polarization beams.  We thank Nathaniel Cuningham and John Bally for sending us the FITS image used to make Figure \ref{orion_hifi_beams}.  We appreciate the input from John Pearson and Harshal Gupta in terms of useful discussions about molecular structure and astrophysics.  Volker Tolls provided valuable information about the WBS noise bandwidth and noise after combining WBS spectral channels.  We have benefited from discussions with  D.  Quan about grain warmup and its impact on molecular cloud chemistry.  The Caltech Submillimeter Observatory is supported by the NSF under award AST-0540882.  This work was carried out in part at the Jet Propulsion Laboratory, California Institute of Technology, under contract with the National Aeronautics and Space Administration.

\begin{center}
\appendix{\textbf{Appendix 1.   HIFI Radiometer Noise Performance}}
\end{center}

These relatively long integrations confirm the excellent performance of the HIFI instrument.  The noise in the spectra of all three \oxy\ spectra is in general quite close to that expected from the radiometer equation.  The major limitation on the precision with which this can be determined is the possible presence of very weak spectral lines and of variable baseline ripple.  We have carried out an extensive analysis of the 774 GHz data, for which the \oxy\ line is accompanied by a moderately large range of frequencies without expected significant lines.  We have data taken in March 2010 and September 2010 with a total of 26,848 s integration time.  The noise (or fluctuation) bandwidth of the channels of the wideband spectrometer (WBS) used to obtain these data is determined by the channel frequency response, which is a quasi-Lorentzian function \citep{tolls2004}.  The resulting noise bandwidth is $\delta\nu_{noise}$ = 1.62 MHz and the nominal channel spacing is 0.57 MHz.   This will vary slightly from the intrinsic spectrometer value due to shifting of channels in the frequency calibration procedure.  

To clarify the comparison between theory and measurements, we have boxcar smoothed the data to a bandwidth of 2 MHz.  According to the simple formula for the noise bandwidth for combination of channels \citep{ossenkopf2008, schieder2001} the noise bandwidth in this case should be close to 3 MHz.  We utilize the single-sideband extracted spectra as described above, giving us a theoretical rms noise for a switched observation $\sigma(T_{MB}) = 4T_{noise}(DSB)\epsilon_{MB}^{-1}/\sqrt{\delta\nu_{noise}t_{int}}$, where $T(DSB)$ is the double sideband noise temperature, $\delta\nu_{noise}$ is the noise bandwidth, and $t_{int}$ is the total (on + off) integration time.  We use for all the data $\epsilon_{MB}$ = 0.75,  $T(DSB)$ = 179 K for the V polarization channel, and $T(DSB)$ = 192 K for the H polarization channel.  The noise temperatures varied by only a few K over the six-month interval spanned by the observations.  

We obtain for the V data $\sigma$ = 0.0062 K, and for the H data $\sigma$ = 0.0053 K, compared to theoretical values 0.0034 K and 0.0036 K, respectively.  The ratio of observed to theoretical noise is 1.82 for V and 1.47 for H, with an average value of 1.65.  These results are impacted by the difficulty of accurately determining the rms from data due to the limited number of ``clean'' channels and time dependent baselines, but they do indicate that the system performance is reasonably close to theoretical.  

The correctness of the noise bandwidth is confirmed by comparison of data smoothed to 4 MHz channel width (5 MHz noise bandwidth) with the 2 MHz channel bandwidth (3 MHz noise bandwidth) data; the ratios for the different data sets have a mean value of 0.76 compared to 0.78 ($\sqrt{3/5}$) expected.  The rms of the 4 MHz smoothed data is 0.0050 K for the V channel and 0.0036 K for the H channel, which average to a factor 1.61 times greater than the theoretical value.  These results suggest that combining spectral channels reduces the noise in a manner close to that theoretically  expected, and that the overall level of fluctuations is a  factor $\sim$ 1.6 greater than that expected strictly from the radiometer equation for integration times up to 2.7$\times$10$^4$ s.

\begin{center}
\appendix{\textbf{Appendix 2.  Effect of Beam Dilution on Line Intensity Ratios}}
\end{center}
\renewcommand{\theequation}{A2-\arabic{equation}}
\setcounter{equation}{0}

 In general, the integrated antenna temperature, $I$, can be written

\begin{equation}
I = \int T_A dv = A_e\frac{A_{ul} N_u hc}{8\pi k} \int P_n(\Omega)C_n(\Omega)d\Omega \lc
\label{intensity_source}
\end{equation}
where in addition to symbols defined following equation \ref{coldens}, we have the antenna effective area $A_e$, the normalized antenna response pattern $P_n$, and the normalized column density distribution, $C_n$.
For an extended uniform ($C_n$ = 1)  source filling the entire antenna pattern, the integral is the antenna solid angle, $\Omega_A$.  If the source is uniform but extended only to fill the main beam, the integral is the main beam solid angle,  $\Omega_{MB}$.  The main beam efficiency $\epsilon_{MB}$ is defined as $\epsilon_{MB} = \Omega_{MB} / \Omega_A$.  

We can still use equation \ref{intensity_source} by considering it to refer to a source that covers at most only the main beam, and a main beam-corrected integrated antenna temperature, $I_{MB} = \epsilon_{MB}^{-1}\int T_A dv$.  From the antenna theorem, $A_e \Omega_A = \lambda^2$, for a given antenna. Then we can rewrite equation \ref{intensity_source} as

\begin{equation}
I_{MB} = \frac{A_{ul} N_u hc \lambda^2}{8\pi k} \frac{1}{\Omega_{MB}}\int_{MB} P_n(\Omega)C_n(\Omega)d\Omega \lc
\label{intensity2}
\end{equation}
where the integral extends over the main beam.   For a uniform source filling the main beam, the integral is just $\Omega_{MB}$ and the integrated main beam temperature, $I_{MB}^{F}$ is equal to the upper level column density multiplied by some constants,

\begin{equation}
I^{src} =  \frac{A_{ul} N_u hc \lambda^2}{8\pi k} \lc
\label{filled}
\end{equation}
where we use the superscript {\it src} to indicate that this intensity is essentially a characteristic of the source itself rather than of the antenna being used for the observations.
For a uniform small source located in the direction of the antenna boresight , $C_n$ = 1 within the extent of the source and = 0 outside.  The integral, which is nonzero over a range that is much smaller than the main beam size so $P_n$ = 1, is equal to the source solid angle, $\Omega_S$.  Thus, the observed intensity, denoted by the superscript {\it obs} is given by

\begin{equation}
I^{obs} = \frac{A_{ul} N_u hc \lambda^2}{8\pi k} \frac{\Omega_S}{\Omega_{MB}} \lp
\label{small}
\end{equation}

The ratio of the observed intensity to that of the source is given by
\begin{equation}
R = I^{obs}/ I^{src} = \Omega_S / \Omega_{MB}  \lp
\label{small_filled}
\end{equation}
Using the antenna theorem again, and defining the aperture efficiency in terms of the physical area of the antenna through $A_e = \epsilon_A A_p$ we find
\begin{equation}R = \frac{\epsilon_A}{\epsilon_{MB}}A_p \Omega_S \lambda^{-2} \lp
\end{equation}
A low frequency (larger wavelength) transition has a small value of $R$, which can be thought of as arising from the small filling factor at the low frequency where the beam size is larger.
To analyze our {\it Herschel} observations of \oxy\ in a reasonably simple manner, we consider a ``small'' source to be one that is smaller than the beam sizes at all three observing frequencies.  
Assuming a constant ratio of aperture to main beam efficiency, which is quite good for {\it Herschel} \citep{olberg2010},  $R$ for two different transitions, 2 and 1, can be written as
\begin{equation}
R(2/1) = [\lambda(1) / \lambda(2)]^2 \lp
\label{small_21}
\end{equation}

Copyright 2011.  All rights reserved.


\begin{thebibliography}{}

\bibitem[Acharyya et al.\ (2007)]{acharyya2007}
Acharyya, K., Fuchs, G.W., Fraser, H.J., van Dishoeck, E.F., \& Linnartz, H. 2007, \aap, 466, 1005

\bibitem[Bally et al.\ (2011)]{bally2011}
Bally, J. Cunningham, N.J., Moeckel, N., Burton, M.G., Smith, N., Frank, A., \& Nordlund, A. 2011, \apj, 727, 113

\bibitem[Beckwith et al.\ (1978)]{beckwith1978}
Beckwith, S., Persson, S.E., Neugebauer, G., \& Becklin, E.E. 1978, \apj, 223, 464

\bibitem[Bergin, Langer, \& Goldsmith (1995)]{bergin1995}
Bergin, E.A., Langer, W.D., \& Goldsmith, P.F. 1995, \apj, 441, 222

\bibitem[Bergin, Melnick, \& Neufeld (1998)]{bergin1998}
Bergin, E.A., Melnick, G.J., \& Neufeld, D.A. 1998, \apj, 499, 777

\bibitem[Bergin et al.\ (2000)]{bergin2000}
Bergin, E.A., Melnick, G.J., Stauffer, J.R., Ashby, M.L.N., Chin, G., Erickson, N.R., Goldsmith, P.F., Harwit, M., Howe, J.E., Kleiner, S.C., Koch, D.G., Neufeld, D.A., Patten, B.M., Plume, R., Schieder, R., Snell, R.L., Tolls, V., Wang, Z., Winnewisser, G., \& Zhang, Y.F. 2000, \apj, 539, L129

\bibitem[Bergman (1995)]{bergman1995}
Bergman, P. 1995, \apj, 445, L167

\bibitem[Black \& Smith (1984)]{black1984}
Black, J. H. \& Smith, P.L. 1984, \apj, 277, 562

\bibitem[Boogert, Blake, \& Tielens (2002)]{boogert2002}
Boogert, A.C.A., Blake, G.A., \& Tielens, A.G.G.M. 2002, \apj, 577, 271

\bibitem[Burton \& Haas (1997)]{burton1997}
Burton, M.G. \& Haas, M.R. 1997, \aap, 327, 309

\bibitem[Carty et al.\ (2006)]{carty2006}
Carty, D., Goddard, A., K\"{o}hler, S.P.K., Sims, I.R., \& Smith, I.W.M. 2006, J. Chem. Phys. A, 110, 3101

\bibitem[Caselli, Hasegawa, \& Herbst (1993)]{caselli1993}
Caselli, P., Hasegawa, T.I., \& Herbst, E. 1993, \apj, 408, 548

\bibitem[Cazaux et al.\ (2010)]{cazaux2010}
Cazaux, S., Cobut, V., Marseille, M., Spaans, M., \& Caselli, P. 2010, \aap, 522, 74

\bibitem[Chi\`{e}ze \& Pineau des For\^{e}ts\ (1989)]{chieze1989}
Chi\`{e}ze J.P. \& Pineau des For\^{e}ts, G. 1989, \aap, 221, 89

\bibitem[Chrysostomou et al.\ (1997)]{chrysostomou1997}
Chrysostomou, A., Burton, M.G., Axon, D.J., Brand, P.W.J.L., Hough, J.H., Bland-Hawthorn, J., \& Geballe, T.R. 1997, \mnras, 289, 605

\bibitem[Collings et al.\ (2004)]{collings2004}
Collings, M.P., Anderson, M.A., Chen, R., Dever, J.W., Viti, S., Williams, D.A., \& McCoustra, M.R.S. 2004, \mnras, 354, 1113

\bibitem[Combes et al.\ (1991)]{combes1991}
Combes, F., Casoli, F., Encrenaz, P., \& Gerin, M. 1991, \aap, 248, 607

\bibitem[Combes \& Wiklind (1995)]{combes1995}
Combes, F. \& Wiklind, T. 1995, \aap, 303, L61

\bibitem[Combes, Wiklind, \& Nakai (1997)]{combes1997}
Combes, F., Wiklind, T., \& Nakai, N. 1997, \aap, 327, L17

\bibitem[Corey, Alexander, \& Schaefer (1986)]{corey1986}
Corey, G.C., Alexander, M.H., \& Schaefer, J. 1986, J. Chem. Phys., 85, 2726

\bibitem[Cuppen et al.\ (2010)]{cuppen2010}
Cuppen, H.M., Ioppolo, S., Romanzin, C., \& Linnartz, H. 2010, Phys. Chem. Chem. Phys., 12, 12077

\bibitem[de Graauw et al.\ (2010)]{degraauw2010}
de Graauw, Th. et al.\ 2010, \aap, 518, L6

\bibitem[Downes et al.\ (1981)]{downes1981}
Downes, D., Genzel, R., Becklin, E.E., \& Wynn-Williams, C.G. 1981, \apj, 244, 869

\bibitem[Erickson et al.\ (1982)]{erickson1982}
Erickson, N.R., Goldsmith, P.F., Snell, R.L., Berson, R.L., Huguenin, G.R., Ulich, B.L., \& Lada, C.J. 1982, \apj, 261, L103

\bibitem[Favre et al.\ (2011)]{favre2011}
Favre, C., Despois, D., Brouillet, N., Baudry, A., Combes, F., Gu\'{e}lin, M., Wootten, A., \& Wlodarczak, G. 2011, arXiv 1103.2548v1

\bibitem[Fraser et al.\ (2001)]{fraser2001}
Fraser, H.J., Collings, M.P., McCoustra, M.R.S., \& Williams, D. A. 2001, \mnras, 327, 1165

\bibitem[Frayer et al.\ (1998)]{frayer1998}
Frayer, D.T., Seaquist, E.R., Thuan, T.X., \& Sievers, A. 1998, \apj, 503, 231

\bibitem[Fuente et al.\ (1993)]{fuente1993}
Fuente, A., Cernicharo, J., Garc\'{i}a-Burillo, S., \& Tejero, J. 1993, \aap, 275, 558


\bibitem[Gezari, Backman, \& Werner (1998)]{gezari1998}
Gezari, D.Y., Backman, D.E., \& Werner, M.W. 1998, \apj, 509, 283

\bibitem[Goicoechea et al.\ (2006)]{goicoechea2006}
Goicoechea, J.R., Cernicharo, J., Lerate, M., Daniel, F., Barlow, M.J., Swinyard, B.M., Lim, T.L., Viti, S., \& Yates, J. 2006, \apj, 641, L49

\bibitem[Goldsmith \& Langer (1978)]{goldsmith1978}
Goldsmith, P.F. \& Langer, W.D. 1978, \apj, 222, 881

\bibitem[Goldsmith et al.\ (1985)]{goldsmith1985}
Goldsmith, P.F., Snell, R.L., Erickson, N.R., Dickman, R.L., Schloerb, F.P., \& Irvine, W.M. 1985, \apj, 289, 613

\bibitem[Goldsmith \& Young (1989)]{goldsmith1989}
Goldsmith, P.F. \& Young, J.S. 1989, \apj, 341, 718

\bibitem[Goldsmith et al.\ (2000)]{goldsmith2000}
Goldsmith, P.F., Melnick, G.J., Bergin, E.A., Howe, J.E., Snell, R.L., Neufeld, D.A., Harwit, M., Ashby, M.L.N., Patten, B.M., Kleiner, S.C., Plume, Rl., Stauffer, J.R., Tolls, V., Wang, Z., Zhang, Y.F., Erickson, N.R., Koch, D.G., Schieder, R., Winnewisser, G., \& Chin, G. 2000, \apj, 539, L123

\bibitem[Goldsmith et al.\ (2002)]{goldsmith2002}
Goldsmith, P.F., Li, D., Bergin, E.A., Melnick, G.J., Tolls, V., Howe, J.E., Snell, R.L., \& Neufeld, D.A. 2001, \apj, 576, 831

\bibitem[Graedel, Langer, \& Frerking (1982)]{graedel1982}
Graedel, T.E., Langer, W.D., \& Frerking, M.A. 1982, \apjs, 48, 321

\bibitem[Gustafsson et al.\ (2003)]{gustafsson2003}
Gustafsson, M., Kristensen, L.E., Cl\'{e}net, Y., Field, D., Lemaire, J.L., Pineau des For\^{e}ts, G., Rouan, D. \& LeCoarer, E. 2003, \aap, 411, 437

\bibitem[Harding et al.\ (2000)]{harding2000}
Harding, L.B., Maergoiz, A.I., Troe, J., \& Ushakov, V.G. 2000, J. Chem. Phys., 113, 11019

\bibitem[Herbst \& Klemperer (1973)]{herbst1973}
Herbst, E. \& Klemperer, W. 1973, \apj, 185, 505

\bibitem[Hirota et al.\ (2007)]{hirota2007}
Hirota, T., Bushimata, T., Choi, Y.K., Honma, M., Imai, H., Iwadate, K., Jike, T., Kameno, S., Kameya, O., Kamohara, R., Kan--Ya, Y., Kawaguchi, N., Kijima, M., Kim, M.K., Kobayashi, H.Kuji, S., Kurayama, T., Manabe, S., Maruyama, K., Matsui, M., Matsumoto, N., Miyaji, T., Nagayama, T., Nakagawa, A., Nakamura, K., Oh, C.S., Omodaka, T., Oyama, T., Sakai, S., Sasao, T., Sato, K., Sato, M., Shibata, K.M., Shintani, M., Tamura, Y., Tsushima, M., \& Yamashita, K., 2007, \pasj, 59, 879 

\bibitem[Hollenbach et al.\ (2009)]{hollenbach2009}
Hollenbach, D., Kaufman, M.J., Bergin,  E.A., \& Melnick, G.J. 2009, \apj, 690, 1497

\bibitem[Howard \& Smith (1981)]{howard1981}
Howard, M.J. \& Smith, I.W.M., J.  Chem. Soc. Faraday Trans. 2, 77, 997

\bibitem[Ioppolo et al.\ (2008)]{ioppolo2008}
Ioppolo, S., Cuppen, H.M, Romanzin, C., van Dishoeck, E.F., \& Linnartz, H. 2008, \apj, 686, 1474

\bibitem[Ioppolo et al.\ (2010)]{ioppolo2010}
Ioppolo, S., Cuppen, H.M., Romanzin, C., van Dishoeck, E.F., \& Linnartz, H. 2010, Phys. Chem. Chem. Phys., 12, 12065

\bibitem[Jensen et al.\ (2000)]{jensen2000}
Jensen, M.J., Bilodeau, R.C., Safvan, C.P., Seiersen, K., Andersen, L.H., Pedersen, H.B., \& Heber, O. 2000, \apj, 543, 764

\bibitem[Kaufman (2010)]{kaufman2010}
Kaufman, M.J. 2010, talk at ``The Stormy Cosmos: The Evolving ISM from Spitzer to Herschel and Beyond,'' November, 2010, available at 
http://www.ipac.caltech.edu/ism2010/sched.shtml

\bibitem[Kim et al.\ (2008)]{kim2008}
Kim, M.K., Hirota, T., Honma, M., Kobayashi, H., Bushimata, T., Choi, Y.K., Imai, H., Iwadate, K., Jike, T., Kameno, S., Kameya, O., Kamohara, R., Kan-Ya, Y., Kawaguchi, N., Kuji, S., Kurayama, T., Manabe, S., Matsui, M., Matsumoto, N., Miyaji, T., Nagayama, T., Nakagawa, A., Oh, C.S., Omodaka, T., Oyama, T., Sakai, S., Sasao, T., Sato, K., Sato, M., Shibata, K.M., Tamura, Y., \& Yamashita, K., 2007, \pasj, 60, 991

\bibitem[Kristensen et al.\ (2003)]{kristensen2003}
Kristensen, L.E., Gustafsson, M., Field, D., Callejo, G., Lemaire, J.L., Vannier, L., \& Pineau des For\^{e}ts, G. 2003, \aap, 412, 727

\bibitem[Lacombe et al.\ (2004)]{lacombe2004}
Lacombe, F., Gendron, E., Rouan, D., Cl\'{e}net, Y., Field, D., Lemaire, J.L, Gustafsson, M., Lagrange, A.-M., Mouillet, D., Rousset, G., Fusco, T., Rousset-Rouvi\`{e}re, L., Servan, B., Marlot, C., \& Feautrier, P. 2004, \aap, 417, L5

\bibitem[Langer \& Graedel (1989)]{langer1989}
Langer, W.D. \& Graedel, T.E. 1989, \apjs, 69, 241

\bibitem[Larsson et al.\ (2007)]{larsson2007}
Larsson, B., Liseau, R., Pagani, L., Bergman, P., Bernath, P., Biver, N., Black, J.H., Booth, R.S., Buat, V., Crovisier, J., Curry, C.L. Dahlgren, M., Encrenaz, P.J., Falgarone, E., Feldman, P.A., Fich, M., Flor\'{e}n, H.G., Fredrixon, M., Frisk, U., Gahm, G.F., Gerin, M., Hagstr\"{o}m, M., Harju, J., Hasegawa, T., Hjalmarson, \AA, Johansson, L.E.B., Justtanont, K., Klotz, A., Kyr\"{o}l\"{a}, E., Kwok, S., Lecacheux, A., Liljestr\"{o}m, T., Llewellyn, E.J., Lundin, S., M\'{e}gie, G., Mitchell, G.F., Murtagh, D., Nordh, L.H., Nyman, L.-\AA., Olberg, M., Olofsson, A.O.H., Olofsson G., Olofsson H., Persson, G., Plume, R., Rickman, H., Ristorcelli, I., Rydbeck, G., Sandqvist, A.A., Sch\'{e}ele, F.V., Serra, G., Torchisky, S., Tothill, N.F., Volk, K., Wiklind, T., Wilson, C.D., Winnberg, A., \& Witt, G. 2007, \aap, 466, 999

\bibitem[Le Bourlot, Pineau des For\^{e}ts, \& Roueff (1995)]{lebourlot1995}
Le Bourlot, J., Pineau des For\^{e}ts, G., \& Roueff, E. 1995, \aap, 297, 251

\bibitem[Leung, Herbst, \& Huebner (1984)]{leung1984}
Leung, C.M., Herbst, E., \& Heubner, W.F. 1984, \apjs, 56, 231

\bibitem[Lique et al.\ (2009a)]{lique2009a}
Lique, F., Jorfi, M., Honvault, P., Halvick, P., Lin, S.Y., Guo, H., Xie, D.Q., Dagdigian, P.J., K\l os, J., \& Alexander, M.H. 2009, J. Chem. Phys., 131, 221104

\bibitem[Lique et al.\ (2009b)]{lique2009b}
Lique, F., van der Tak, F.F.S., K\l os, J., Bulthuis, J., \& Alexander, M. 2009, \aap, 493, 557

\bibitem[Lique (2010)]{lique2010}
Lique, F. 2010, J. Chem. Phys., 132, 044311

\bibitem[Liseau et al.\ (2010)]{liseau2010}
Liseau, R., Larsson, B., Bergman, P., Pagani, L., Black, J.H., Hjalmarson, \AA., \& Justtanont, K. 2010, \aap, 510, A98

\bibitem[Liszt \& Vanden Bout (1985)]{liszt1985}
Liszt, H.S. \& Vanden Bout, P.A. 1985, \apj, 291, 178

\bibitem[Liszt (1992)]{liszt1992}
Liszt, H.S. 1992, \apj, 386, L139

\bibitem[Luhman et al.\ (1994)]{luhman1994}
Luhman, M.L., Jaffe, D.T., Keller, L.D., \& Pak, S. 1994, \apj, 436, 185L

\bibitem[Mar\'{e}chal et al.\ (1997a)]{marechal1997a}
Mar\'{e}chal, P., Pagani, L., Langer, W.D., \& Castets, A. 1997, \aap, 318, 252

\bibitem[Mar\'{e}chal et al.\ (1997b)]{marechal1997b}
Mar\'{e}chal, P., Viala, Y.P., \& Benayoun, J.J. 1997, \aap, 324, 221

\bibitem[Mar\'{e}chal et al.\ (1997c)]{marechal1997c}
Mar\'{e}chal, P., Viala, Y.P., \& Pagani, L. 1997, \aap, 328, 617

\bibitem[Masson \& Mundy (1988)]{masson1988}
Masson, C.R. \& Mundy, L.G. 1988, \apj, 324, 543

\bibitem[Melnick et al.\ (2000)]{melnick2000}
Melnick, G.J., Stauffer, J.R., Ashby, M.L.N., Bergin, E.A., Chin, G., Erickson, N.R., Goldsmith, P.F., Harwit, M., Howe, J.E., Kleiner, S.C., Koch, D.G., Neufeld, D.A., Patten, B.M., Plume, R., Schieder, R., Snell, R.L., Tolls, V., Wang, Z., Winnewisser, G., \& Zhang, Y.F. 2000, \apj, 539, L77


\bibitem[Menten et al.\ (2007)]{menten2007}
Menten, K.M., Reid, M.J., Forbrich, J., \& Brunthaler, A. 2007, \aap, 474, 515

\bibitem[Miyauchi et al.\ (2008)]{miyauchi2008}
Miyauchi, N., Hidaka, H., Chigai, T., Nagaoka, A., Watanabe, N., \& Kouchi, A. 2008, Chem. Phys. Lett. 456, 27

\bibitem[Mul et al.\ (1983)]{mul1983}
Mul, P.M., McGowan, J.W., Defrance, P., \& Mitchell, J.B.A. 1983, J. Phys. B, 16, 2099

\bibitem[M\"{u}ller et al.\ (2001)]{mueller2001}
M\"{u}ller, H.S.P., Thorwith, S., Roth, D.A., \& Winnewisser, G. 2001, \aap, 370, L49

\bibitem[Murata et al.\ (1991)]{murata1991}
Murata, Y., Kawabe, R., Ishiguro, M., Hasegawa, T., \& Hayashi, M. 1991, in IAU Symposium 147, Fragmentation of Molecular Clouds and Star formation, E. Falgarone, F. Boulanger, \& G. Duvert (eds).  Dordrecht:  Kluwer, 357

\bibitem[Murata et al.\ (1992)]{murata1992}
Murata, Y., Kawbe, R., Ishiguro, M., Morita, K.-I., Hasegawa, T., \& Hayashi, M. 1992, \pasj, 44, 381

\bibitem[Neau et al.\ (2000)]{neau2000}
Neau, A., Al Khalili, A., Ros\'{e}n, S., Le Padellec, A., Derkatch, A.M., Shi, W., Vikor, L., Semaniak, J., Thomas, R., N\aa g\aa rd, M.B., Andersson, K., Danared, H., \& af Ugglas, M. 2000, J. Chem. Phys., 11, 1762

\bibitem[Nenkova, Ivezi\'{c}, \& Elitzur (2000)]{nenkova2000}
Nenkova, M.,Ivezi\'{c}, \v{Z}, \& Elitzur, M. 2000, Proc. Conf. Thermal Emission Spectroscopy and Analysis of Dust, Disks, and Regoliths, M. Sitko, A.Sprague, \& D. Lynch, eds.  ASP Conf. Series, 196, 77

\bibitem[Nordh et al.\ (2003)]{nordh2003}
Noordh, H.L., von Sch\'{e}ele, F., Frisk, U., Ahola, K., Booth, R.S., Encrenaz, P.J., Hjalmarson, \AA., Kendall, D., Kyr\"{o}l\"{a}, E., Kwok, S., Lecacheux, A., Leppelmeier, G., Llewellyn, E.J., Mattila, K., M\'{e}gie, G., Murtagh, D., Rougeron, M., \& Witt, G. 2003, \aap, 402, L21

\bibitem[\"{O}berg, van Dishoeck, \& Linnartz (2009)]{oberg2009a}
\"{O}berg, K.I., van Dishoeck, E.F., \& Linnartz, H. 2009, \aap, 496, 281

\bibitem[\"{O}berg et al.\ (2009)]{oberg2009b}
\"{O}berg, K.I., Linnartz, H., Visser, R., \& van Dishoeck, E.F. 2009, \apj, 603, 1209


\bibitem[Olberg (2010)]{olberg2010}
Olberg, M. 2010, HIFI ICC Tech. Note ICC/2010-nnn, issue 1.1

\bibitem[Olofsson et al.\ (1998)]{olofsson1998}
Olofsson, G., Pagani, L, Tauber, J., Febvre, P., Deschamps, A., Encrenaz, P., Flor\'{e}n, H.-G., George, S., Lecomte, B., Ljung, B., Nordh, L., Pardo, J.R., Peron, I. Sj\"{o}kvist, M., Stegner, L., Stenmark, L. \& Ullberg, C. 1998, \aap, 339, L81

\bibitem[Orlikowski (1985)]{orlikowski1985}
Orlikowski, T. 1985, Molec. Phys., 56, 35

\bibitem[Ossenkopf (2008)]{ossenkopf2008}
Ossenkopf, V. 2008, \aap, 479, 215

\bibitem[Ott (2010)]{ott2010}
Ott, S. 2010 in Astronomical Data Analysis Software and Systems XIX, eds. Y. Mizuno, K. I. Morita, \& M. Ohishi, ASP Conf. Ser. 434, 139

\bibitem[Pagani, Langer, \& Castets (1993)]{pagani1993}
Pagani, L., Langer, W.D., \& Castets, A. 1993, \aap, 274, L13

\bibitem[Pagani et al.\ (2003)]{pagani2003}
Pagani, L., Olofsson, A.O.H., Bergman, P., Bernath, P., Black, J.H., Booth, R.S., Buat, V., Crovisier, J., Curry, C.L., Encrenaz, P.J., Falgarone, E., Feldman, P.A., Fich, M., Gloren, H.G., Frisk, U., Gerin, M., Gregersen, E.M., Harju, J., Hasegawa, T., Hjalmarson, \AA., Johansson, L.E.B., Kwok, S., Larsson, B., Lecacheux, A., Liljestr\"{o}m, T., Lindqvist, M., Liseau, R., Mattila, K., Mitchell, G.F., Nordh, L.H., Olberg, M., Olofsson, G., Ristorcelli, I., Sandqvist, \AA., von Scheele, F., Serra, G., Tothill, N.F., Volk, K., Wiklind, T., \& Wilson, C.D. 2003 \aap, 402, L77

\bibitem[Parmar, Lacy, \& Achtermann (1994)]{parmar1994}
Parmar, P.S., Lacy, J.H., \& Achtermann, J.M. 1994, \apj, 430, 786

\bibitem[Pardo, Cernicharo, \& Phillips (2005)]{pardo2005}
Pardo, J.R., Cernicharo, J., \& Phillips, T.G. 2005, \apj, 634, L31

\bibitem[Pauls et al.\ (1983)]{pauls1983}
Pauls, T.A., Wilson, T.L., Bieging, J.H., \& Martin, R.N. 1983, \aap, 124, 23

\bibitem[Penzias (1981)]{penzias1981}
Penzias, A.A. 1981, \apj, 249, 518

\bibitem[Pilbratt et al.\ (2010)]{pilbratt2010}
Pilbratt, G.L. et al.\ 2010, \aap, 518, L1

\bibitem[Plambeck \& Wright (1987)]{plambeck1987}
Plambeck, R.L. \& Wright, M.C.H. 1987, \apj, 317, L101

\bibitem[Pontoppidan et al.\ (2003)]{pontoppidan2003}
Pontoppidan, K.M., Fraser, H.J., Dartois, E., Thi, W.--F., van Dishoeck, E.F., Boogert, A.C.A., D'Hendecourt, L., Tielens, A.G.G.M., \& Bisschop, S.E. 2003, \aap, 408, 981

\bibitem[Prasad \& Huntress (1980)]{prasad1980}
Prasad, S.S. \& Huntress, W.T. Jr. 1980, \apjs, 43, 1



\bibitem[Qu\'{e}mener, Balakrishnan, \& Kendrick (2008)]{quemener2008}
Qu\'{e}mener, G., Balakrishnan, N., \& Kendrick, B.K. 2008, J. Chem. Phys., 129, 224309

\bibitem[Rosenthal, Bertoldi, \& Drapatz (2000)]{rosenthal2000}
Rosenthal, D., Bertoldi, F., \& Drapatz, S. 2000, \aap, 356, 705

\bibitem[Salas et al.\ (1999)]{salas1999}
Salas, L. Rosado, M., Cruz-Gonz\'{a}lez, I., Guti\'{e}rrez, L., Valdez, J., Bernal, A., Luna, E., Ruiz, E., \& Lazo, F. 1999, \apj, 511, 822

\bibitem[Sandqvist et al.\ (2008)]{sandqvist2008}
Sandqvist, Aa., Larsson, B., Hjalmarson, \AA, Bergman, P., Bernath, P., Frisk, U., Olberg, M., Pagani, L., \& Ziurys, L. 2008, \aap, 482, 849

\bibitem[Sandstrom et al.\ (2007)]{sandstrom2007}
Sandstrom, K.M., Peek, J.E.G., Bower, G.C., Bolatto, A.D., \& Plambeck, R.L. 2007, \apj, 667, 1161

\bibitem[Schieder \& Kramer (2001)]{schieder2001}
Schieder, R. \& Cramer, C. 2001, \aap, 373, 746

\bibitem[Schild, Miller, \& Tennyson (1997)]{schild1997}
Schild, H., Miller, S., \& Tennyson, J. 1997, \aap, 318, 608

\bibitem[Sch\"{o}ier et al.\ (2005)]{schoier2005}
Sch\"{o}ier, F.L., van der Tak, F.F.S., van Dishoeck, E.F., \& Black, J.H. 2005, \aap, 432, 369

\bibitem[Schultz et al.\ (1999)]{schultz1999}
Shultz, A.S.B., Colgan, S.W.J., Erickson, E.F., Kaufman, M.J., Hollenbach, D.J., O'Dell, C.R., Young, E.T., \& Chen, H. 1999, \apj, 511, 282

\bibitem[Scoville et al.\ (1982)]{scoville1982}
Scoville, N.Z., Hall, D.N.B., Kleinmann, S.G., \& Ridgway, S.T. 1982, \apj, 253, 136

\bibitem[Smith et al.\ (1984)]{smith1984}
Smith, P.L., Griesinger, H.E., Black, J.H., Yoshino, K., \& Freeman, D.E. 1984, \apj, 277, 569

\bibitem[Smith \& Stewart (1994)]{smith1994}
Smith, I.W.M., \& Stewart, D.W.A. 1991, J. Chem. Soc. Faraday Trans. 2, 90, 3221

\bibitem[Stolovy et al.\ (1998)]{stolovy1998}
Stolovy, S., Burton, M.G., Erickson, E.F., Kaufman, M.J., Chrysostomou, A., Young, E.T., Colgan, S.W.J., Axon, D.J., Thompson, R.I., Rieke, M.J., \& Schneider, G. 1994, \apj, 492, L151

\bibitem[Tolls et al.\ (2004)]{tolls2004}
Tolls, V., Melnick, G.J., Ashby, M.L.N., Bergin, E.A., Gurwell, M.A., Kleiner, S.C., Patten, B.M., Plume, R., Stauffer, J.R., Wang, Z., Zhang, Y.F., Chin, G., Erickson, N.R., Snell, R.L., Goldsmith, P.F., Neufeld, D.A. , Schieder, R., \& Winnewisser, G. 2004, \apjs, 152, 137

\bibitem[Valkenberg (1995)]{valkenberg1995}
Valkenberg, R. 1995, Diplom. in Physik thesis, Rheinisch--Westf\"{a}lischen Tech. Hochschule, Aachen

\bibitem[Vandenbussche et al.\ (1999)]{vandenbussche1999}
Vandenbussche, B., Ehrenfreund, P., Boogert, A.C.A., van Dishoeck, E.F., Schutte, W.A., Gerakines, P.A., Chiar, J., Tielens, A.G.G.M., Keane, J., Whittet, D.C.B., Breitfellner, M., \& Burgdorf, M. 1999, \aap, 346, L57

\bibitem[Van der Tak et al.\ (2007)]{vandertak2007}
Van der Tak, F.F.S., Black, J.H., Sch\"{o}ier, F.L., Jansen, D.J., \& van Dishoeck, E.F. 2007, \aap, 468, 627

\bibitem[Vannier et al.\ (2001)]{vannier2001}
Vannier, L., Lemaire, J.L. , Field, D., Pineau des For\^{e}ts, G., Pijpers, F.P., \& Rouan, D. 2001, \aap, 366, 651

\bibitem[Vejby-Christensen et al.\ (1997)]{vejby1997}
Vejby-Christensen, L, Andersen, L.H., Heber, O., Kella, D., Pedersen, H.B., Schmidt, H.T., \& Zajfman, D. 1997, \apj, 483, 531

\bibitem[Wakelam et al.\ (2005)]{wakelam2005}
Wakelam, V., Selsis, F., Herbst, E., \& Caselli, P. 2005, \aap, 444, 883

\bibitem[Whittet (2010)]{whittet2010}
Whittet, D.C.B. 2010, \apj, 710, 1009

\bibitem[Wilson et al.\ (2000)]{wilson2000}
Wilson, T.L., Gaume, R.A., Gensheimer, P., \& Johnston, K.J.  2000, \apj, 538, 665

\bibitem[Wilson et al.\ (2005)]{wilson2005}
Wilson, C.D., Olofsson, A.O.H., Pagani, L., Booth, R.S., Frisk, U., Hjalmarson, \AA., Olberg, M., \& Sandqvist, Aa. 2005, \aap, 443, L5

\bibitem[Wright et al.\ (1992)]{wright1992}
Wright, M.C.H., Sandell, G., Wilner, D.J., \& Plambeck, R.L. 1992, \apj, 393, 225

\bibitem[Wright, Plambeck, \& Wilner (1996)]{wright1996}
Wright, M.C.H, Plambeck, R.L., \& Wilner, D.J. 1998, \apj, 469, 216

\bibitem[Xie, Allen, \& Langer (1995)]{xie1995}
Xie, T., Allen, M., \& Langer, W.D. 1995, \apj, 440, 674

\bibitem[Xu et al.\ (2007)]{xu2007}
Xu, C., Xie, D., Honvault, P., Lin, S.Y., \& Guo, H. 2007, J. Chem. Phys., 127, 024304

\end{thebibliography}
\end{document}